\documentclass[english,notitlepage,nofootinbib]{revtex4-1}
\usepackage[T1]{fontenc}
\usepackage[latin9]{inputenc}
\usepackage{babel}
\usepackage{amsmath}
\usepackage{amssymb}
\usepackage{graphicx}
\PassOptionsToPackage{normalem}{ulem}
\usepackage{ulem}
\usepackage[bookmarks=false,
 breaklinks=false,pdfborder={0 0 1},backref=false,colorlinks=false]
 {hyperref}

\makeatletter

\usepackage{babel}

\makeatother

\begin{document}
\title{Exclusive photoproduction of $\eta_{c}\gamma$ pairs with large invariant mass}
\author{Marat Siddikov }
\affiliation{Departamento de Física, Universidad Técnica Federico Santa María,~~~~\\
 y Centro Científico - Tecnológico de Valparaíso, Casilla 110-V, Valparaíso,
Chile}
\begin{abstract}
In this preprint we analyzed the exclusive photoproduction of $\eta_{c}\gamma$ pairs
in the collinear factorization framework and evaluated its cross-section
in leading order in the strong coupling $\alpha_{s}$.
We found that this process is mostly sensitive to the behavior of the unpolarized
gluon GPD $H_{g}$ in the Efremov-Radyushkin-Brodsky-Lepage kinematics.
We estimated numerically the cross-section and the expected counting
rates in the kinematics of middle-energy photoproduction experiments
which can be realized at the future Electron-Ion Collider and in
the kinematics of ultraperipheral collisions at LHC. We found that
the cross-section is sufficiently large for dedicated experimental
study. 
\end{abstract}
\maketitle

\section{Introduction}

The Generalized Parton Distributions (GPDs) encode information about
the nonperturbative dynamics of different parton flavors in the hadronic
target, and for this reason have been intensively studied both theoretically
and experimentally in the last two decades~\cite{Diehl:2000xz,Goeke:2001tz,Diehl:2003ny,Guidal:2013rya,Boer:2011fh,Dutrieux:2021nlz,Burkert:2022hjz,AbdulKhalek:2021gbh,Accardi:2012qut}.
At present, it is not possible to evaluate the GPDs from the first
principles, and for this reason studies of these objects rely on phenomenological
extractions from experimental data. Most of the currently available
phenomenological parametrizations of the GPDs are based on experimental
studies of $2\to2$ processes, like Deeply Virtual Compton Scattering
(DVCS) and Deeply Virtual Meson Production (DVMP). However, as was
discussed in~~\cite{Kumericki:2016ehc,Bertone:2021yyz,Moffat:2023svr},
these channels do not fix uniquely the GPDs due to poor sensitivity
of the physical observables to behavior of the GPDs in some kinematic
regions (the so-called ``\emph{shadow GPD problem}''). While the
experimental data allow to \emph{constrain} reasonably well some of
the GPDs, at present these constraints are very loose in the Efremov-Radyushkin-Brodsky-Lepage
(ERBL) kinematics, for the transversity GPDs, and the GPDs of gluons
in general~\cite{Guo:2023ahv}. For this reason, the interest of
the community has shifted towards search of new channels which could
be used for complementary studies, and various $2\to3$ processes
have been suggested as novel tools for studies of the GPDs~\cite{GPD2x3:9,GPD2x3:8,GPD2x3:7,GPD2x3:6,GPD2x3:5,GPD2x3:4,GPD2x3:3,GPD2x3:2,GPD2x3:1,Duplancic:2022wqn,Qiu:2024mny,Qiu:2023mrm,Deja:2023ahc,Siddikov:2022bku,Siddikov:2023qbd}.
Due to sensitivity of each process to unique combination of different
GPD flavors in different kinematic domains, all these channels naturally
complement each other, and the quest for new channels remains open.
As was discussed in~~\cite{GPD2x3:10,GPD2x3:11}, the factorization
theorems for such processes in general require presence of hard scale
and good kinematic separation of the produced hadrons. Most of the
above-mentioned studies focused on production of light meson or photon
pairs, whose amplitudes are dominated by quark GPDs (both in chiral
odd and chiral even sectors). The gluon GPDs also may contribute in
some of these channels, however usually these contributions are commingled with contribution of quarks, complicating the phenomenological analysis.
Furthermore, recently it was discovered in~\cite{Duplancic:2024opc,Nabeebaccus:2023}
that the gluonic coefficient function of the processes like $\gamma p\to\pi^{0}\gamma p$
suffers from nontrivial overlapping singularities, and the applicability
of the collinear factorization is justified only if the distribution
amplitude of produced light meson vanishes rapidly enough at the endpoints.

In order to take advantage of the $2\to3$ processes for analysis
of the gluon GPDs, we suggest to study the production of heavy quarkonia-photon
pairs. Due to heavy mass of the quarkonia, which plays the role of
natural hard scale, it is possible to use the perturbative methods
even in the photoproduction regime~\cite{Korner:1991kf,Neubert:1993mb},
and avoid many soft and collinear singularities which may occur in
diagrams with massless quarks. The hadronization of $\bar{Q}Q$ pair
into final state quarkonium is described in NRQCD framework, which
allows to incorporate systematically various perturbative corrections~\cite{Bodwin:1994jh,Maltoni:1997pt,Brambilla:2008zg,Feng:2015cba,Brambilla:2010cs,Cho:1995ce,Cho:1995vh,Baranov:2002cf,Baranov:2007dw,Baranov:2011ib,Baranov:2016clx,Baranov:2015laa}.
The single-quarkonium production $\gamma p\to J/\psi\,p$ has been
studied in detail both theoretically and experimentally~\cite{DVMPcc1,DVMPcc2,DVMPcc3,DVMPcc4},
however it shares all the above-mentioned limitations of DVCS and
DVMP. The process $\gamma p\to J/\psi\,\gamma p$ requires $C$-odd
exchanges in the $t$-channel, and for this reason is strongly suppressed.
However, the process $\gamma p\to\eta_{c}\,\gamma p$ presents an
interesting opportunity for study of the gluon GPDs. Previously this
channel has been studied in~\cite{HarlandLang:2018ytk}, assuming
that it proceeds via a radiative decay of higher excited states (mainly
$\gamma p\to J/\psi p\to\eta_{c}\gamma\,p$). Though this mechanism
has a relatively large cross-section, it is controlled by the $J/\psi$
photoproduction cross-section and thus does not bring any new information
for studies of GPDs. Fortunately, it is possible to eliminate this
contribution imposing cuts on invariant mass of $\eta_{c}\gamma$
pair. In this paper, we will focus on production of $\eta_{c}\gamma$
pairs with large invariant mass, $M_{\gamma\eta_{c}}\ge3.5-4\,{\rm GeV}^{2}$,
where the feed-down contributions are negligible. This kinematic regime
may be studied in low- and middle-energy photon-proton collisions
in ultraperipheral kinematics at LHC, as well as electron-proton collisions
at the future Electron Ion Collider (EIC)~\cite{Accardi:2012qut,AbdulKhalek:2021gbh,Burkert:2022hjz}
and possible experiments at JLAB after 22 GeV upgrade~\cite{Accardi:2023chb}. 

The suggested $\eta_{c}\gamma$ pair photoproduction (with undetected
photon) is also interesting as a potential background to the $\eta_{c}$
photoproduction $\gamma p\to\eta_{c}p.$ The latter subprocess conventionally
has been considered as one of the most promising channels for studies
of odderons~\cite{Odd4,Odd5,Odd6,Odd7}, and significant efforts
have been dedicated to its theoretical understanding and experimental detection.
The $\eta_{c}\gamma$ pair photoproduction does not require $C$-odd
exchanges in $t$-channel, and for this reason could constitute a
sizable background which sets the limits on detectability of odderons
via $\eta_{c}$ photoproduction at future experiments. 

The paper is structured as follows. Below in Section~\ref{sec:Formalism}
we discuss in detail the kinematics of the process and the framework
for evaluation of the amplitude of the process.  In Section~\ref{sec:Numer}
we present our numerical estimates for cross-sections and counting
rates of the process. In subsection~\ref{subsec:odderon} we compare
the photoproduction cross-sections for $\eta_{c}$ and $\eta_{c}\gamma$
with integrated out photon. Finally, in Section~\ref{sec:Conclusions}
we draw conclusions.

\section{Theoretical framework}

\label{sec:Formalism} For analysis of the $\eta_{c}\gamma$ -photoproduction
we may use the framework developed in~\cite{GPD2x3:8,GPD2x3:9} for
exclusive photoproduction of\emph{ light} meson-photon pairs $\gamma\pi,\,\gamma\rho$
with large invariant mass. The extension of those results to $\eta_{c}\gamma$
photoproduction is straightforward, however we can no longer disregard
the mass of the meson $M_{\eta_{c}}$. It is more appropriate to consider
$M_{\eta_{c}}$ as a hard scale on par with the invariant mass $M_{\gamma\eta_{c}}$,
keeping it both in the kinematic relations and in coefficient functions.
Both in electroproduction and ultraperipheral hadroproduction the
spectrum of equivalent photons is dominated by quasi-real photons,
so we'll focus on the photoproduction by transversely polarized photons
with zero virtuality $Q=0$. In the following subsections~\ref{subsec:Kinematics},
~\ref{subsec:Amplitudes} we briefly introduce the main kinematic
variables used for description of the process and discuss the evaluation
of the amplitudes in the collinear factorization approach. 

\subsection{Kinematics of the process}

\label{subsec:Kinematics} In what follows, we will perform evaluations
in the photon-proton collision frame, where the photon and the incoming
proton move in the direction of axis $z$. We will use notations $q$
for the momentum of the incoming photon, $P_{{\rm in}},P_{{\rm out}}$
for the momentum of the proton before and after interaction, $k$
for the momentum of the emitted (outgoing) photon, and $p_{\eta_{c}}$
for the momentum of produced $\eta_{c}$ meson. In this frame the
light-cone decomposition of the particles' momenta may be written
as~\cite{GPD2x3:8,GPD2x3:9} 
\begin{align}
q^{\mu} & =n^{\mu}\label{eq:q}\\
P_{{\rm in}}^{\mu} & =\left(1+\xi\right)p^{\mu}+\frac{m_{N}^{2}}{s(1+\xi)}n^{\mu},\\
P_{{\rm out}}^{\mu} & =\left(1-\xi\right)p^{\mu}+\frac{m_{N}^{2}+\Delta_{\perp}^{2}}{s(1+\xi)}n^{\mu}+\Delta_{\perp}^{\mu},\\
p_{\eta_{c}}^{\mu} & =\alpha_{\eta_{c}}n^{\mu}+\frac{\left(\boldsymbol{p}_{\perp}+\boldsymbol{\Delta}_{\perp}/2\right)^{2}+M_{\eta_{c}}^{2}}{\alpha_{\eta_{c}}s}p^{\mu}-\boldsymbol{p}_{\perp}-\frac{\boldsymbol{\Delta}_{\perp}}{2},\\
k^{\mu} & =\left(1-\alpha_{\eta_{c}}\right)n^{\mu}+\frac{\left(\boldsymbol{p}_{\perp}-\boldsymbol{\Delta}_{\perp}/2\right)^{2}}{\left(1-\alpha_{\eta_{c}}\right)s}p^{\mu}+\boldsymbol{p}_{\perp}-\frac{\boldsymbol{\Delta}_{\perp}}{2},\label{eq:k}
\end{align}
where the basis light-cone vectors $p^{\mu},\,n^{\mu}$ are defined
as
\begin{equation}
p^{\mu}=\frac{\sqrt{s}}{2}\left(1,0,0,1\right),\qquad n^{\mu}=\frac{\sqrt{s}}{2}\left(1,0,0,-1\right),\qquad p\cdot n=\frac{s}{2}.
\end{equation}
In what follows we will use the invariant Mandelstam variables
\begin{align}
S_{\gamma N} & \equiv W^{2}=\left(q+P_{{\rm in}}\right)^{2}=s\left(1+\xi\right)+m_{N}^{2},\\
t & =\left(P_{{\rm out}}-P_{{\rm in}}\right)^{2}=-\frac{1+\xi}{1-\xi}\Delta_{\perp}^{2}-\frac{4\xi^{2}m_{N}^{2}}{1-\xi^{2}}.\label{eq:tDep}
\end{align}
From Eq.~(\ref{eq:tDep}) we can see that at given $\xi$, the invariant
momentum transfer $t$ is bound by 
\[
t\le t_{{\rm min}}=-\frac{4\xi^{2}m_{N}^{2}}{1-\xi^{2}}.
\]
In what follows we will also use the variables
\begin{align}
 & u'=\left(p_{\eta_{c}}-q\right)^{2},\qquad t'=\left(k-q\right)^{2},\qquad M_{\gamma\eta_{c}}^{2}=\left(k+p_{\eta_{c}}\right)^{2}\label{eq:uPrimetPrime}
\end{align}
which are related as
\begin{align}
 & -u'-t'=M_{\gamma\eta_{c}}^{2}-M_{\eta_{c}}^{2}-t.\label{eq:Constr}
\end{align}
The polarization vector of the incoming/outgoing real photon
with momentum $\boldsymbol{k}$ and helicity $\lambda$ in the light-cone gauge is given by
\begin{equation}
\varepsilon_{T}^{(\lambda=\pm1)}(\boldsymbol{k})=\left(0,\,\frac{\boldsymbol{e}_{\lambda}^{\perp}\cdot\boldsymbol{k}_{\perp}}{k^{+}},\boldsymbol{e}_{\lambda}^{\perp}\right),\qquad\boldsymbol{e}_{\lambda}^{\perp}=\frac{1}{\sqrt{2}}\left(\begin{array}{c}
1\\
i\lambda
\end{array}\right).\label{eq:PolVector}
\end{equation}

\begin{figure}

\includegraphics[scale=0.4,height=8.5cm]{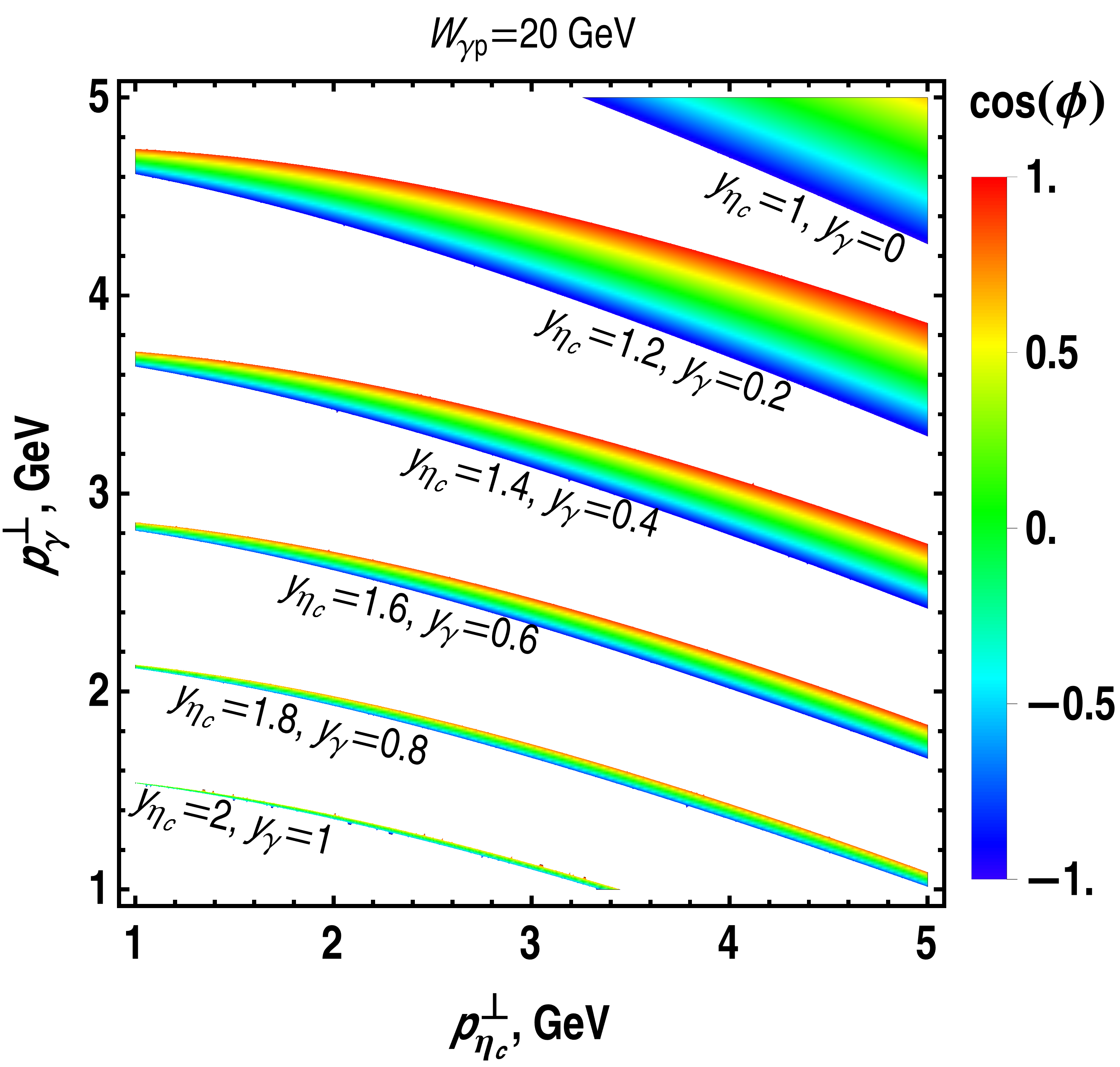}\includegraphics[scale=0.4,height=8.5cm]{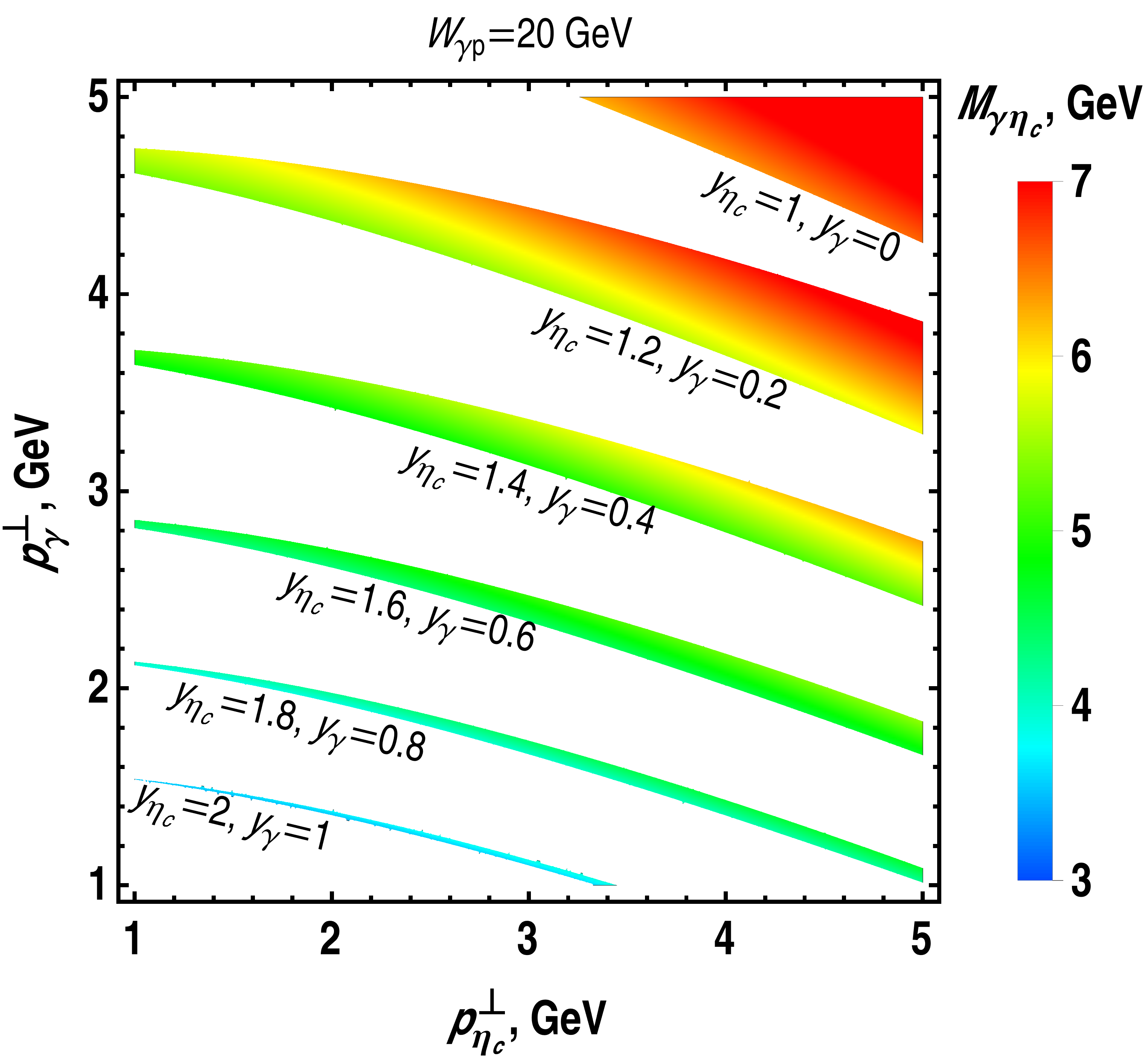}

\caption{\protect\label{fig:Domain}(Color online) Colored bands represent
kinematically allowed region for $\eta_{c}\gamma$ pair production
\uline{at fixed invariant energy} $W$ and fixed rapidities $y_{\eta_{c}},\,y_{\gamma}$.
For both particles we chose positive sign of rapidity in direction
of the photon. Since photon moves in \textquotedblleft minus\textquotedblright{}
direction in~(\ref{eq:q}), this corresponds to nonstandard definition
of rapidity $y_{a}=\frac{1}{2}\ln\left(k_{a}^{-}/k_{a}^{+}\right),\qquad a=\eta_{c},\gamma$.
The increase of rapidities implies increase of the longitudinal momenta,
and thus in view of energy conservation corresponds to smaller transverse
momenta. In the left plot the color of each pixel reflects the azimuthal
angle $\phi$ between the transverse momenta $\boldsymbol{p}_{\eta_{c}}^{\perp}=-\boldsymbol{p}_{\perp}-\boldsymbol{\Delta}_{\perp}/2$
and $\boldsymbol{p}_{\gamma}^{\perp}=\boldsymbol{p}_{\perp}-\boldsymbol{\Delta}_{\perp}/2$.
Similarly, in the right plot the color reflects the value of invariant
mass $M_{\gamma\eta_{c}}$ in the chosen kinematics.}
\end{figure}

\begin{figure}
\includegraphics[width=6cm]{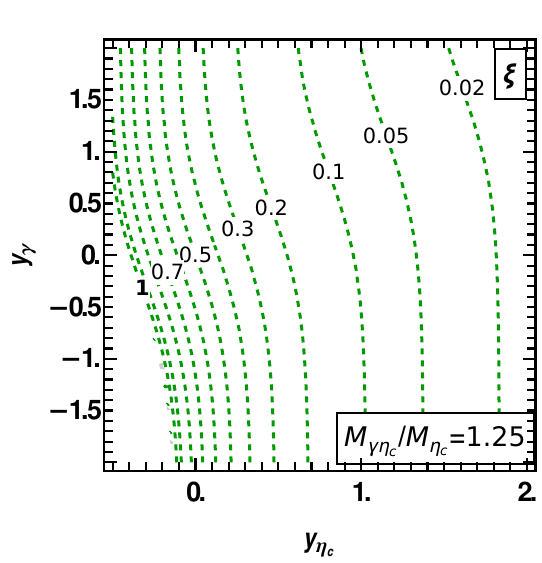}\includegraphics[width=6cm]{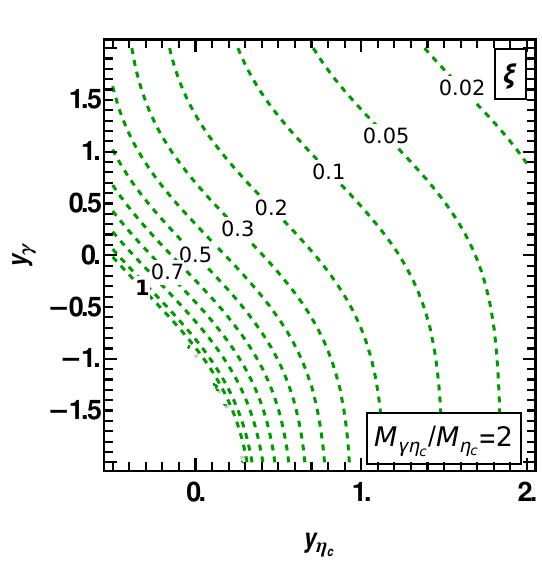}\includegraphics[width=6cm]{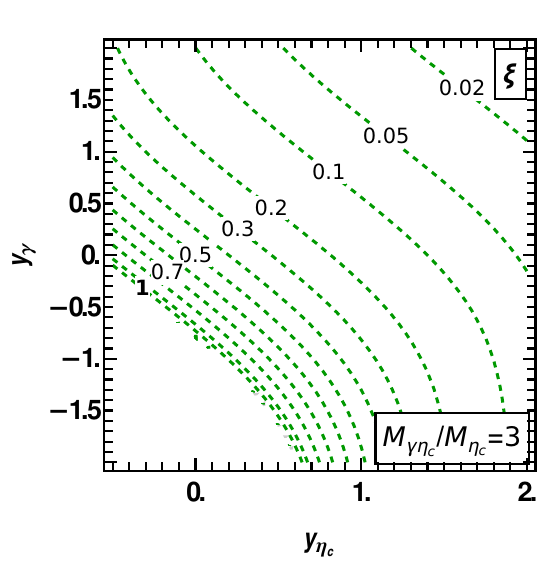}

\includegraphics[width=6cm]{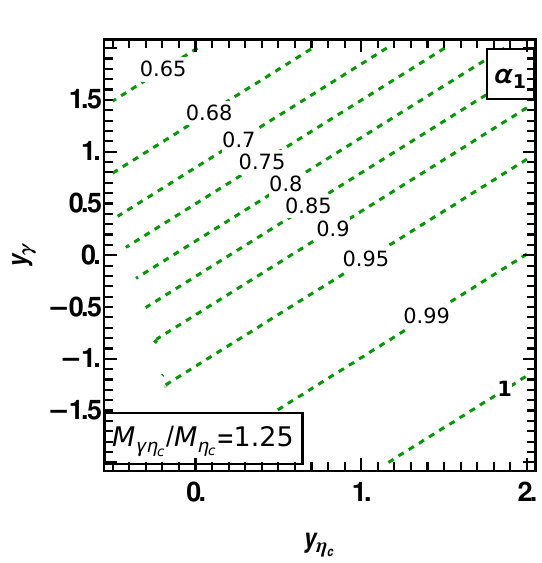}\includegraphics[width=6cm]{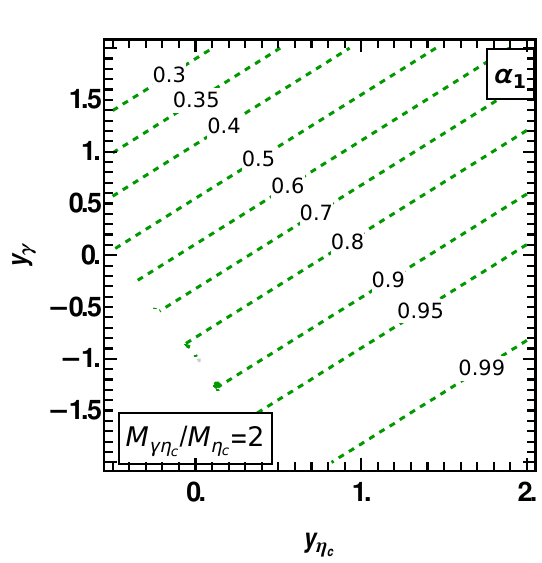}\includegraphics[width=6cm]{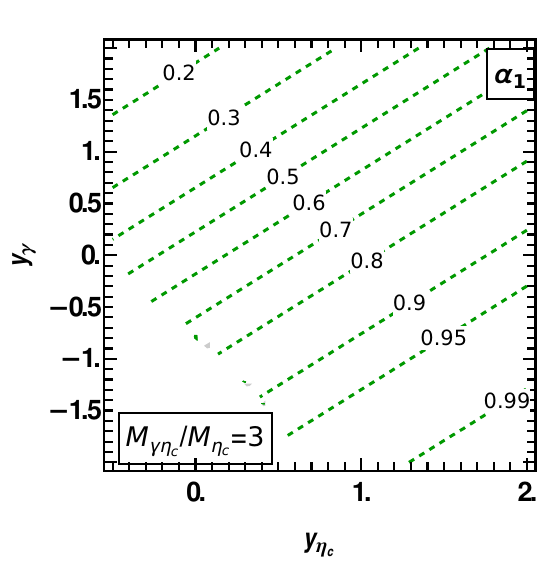}\caption{\protect\label{fig:xiRapidity}The contour plot illustrates the relation
of the light-cone variable $\xi,\,\alpha_{1}\equiv\alpha_{\eta_{c}}$
with the rapidities of final state particles $y_{\eta_{c}},y_{\gamma}$
for different values of invariant mass $M_{\gamma\eta_{c}}$. For
both particles we choose positive sign of rapidity for motion in direction
of the incoming photon. For the sake of simplicity we consider the
kinematics with zero transverse momenta of $\gamma,\eta_{c}$, which
gives the dominant contribution in the total cross-section. The labels
on contour lines show the values of $\xi,\,\alpha_{\eta_{c}}$.}
\end{figure}

The parametrization~~(\ref{eq:q}-\ref{eq:k}) implicitly implements
various kinematic constraints on momenta of the produced particles
which follow from onshellness of final state particles and energy-momentum
conservation. In order to understand better these constraints, in
the Figure~(\ref{fig:Domain}) we have shown the kinematically allowed
domains using conventional rapidities and transverse momenta of the
final-state particles. The plot clearly shows that for a given energy
$W$ and rapidities $y_{\eta_{c}},y_{\gamma}$, the transverse momenta
$\boldsymbol{p}_{\eta_{c}}^{\perp}=-\boldsymbol{p}_{\perp}-\boldsymbol{\Delta}_{\perp}/2$
and $\boldsymbol{p}_{\gamma}^{\perp}=\boldsymbol{p}_{\perp}-\boldsymbol{\Delta}_{\perp}/2$
may take values only in a limited range. The color of each pixel in
the Figure~~(\ref{fig:Domain}) illustrates the value of the azimuthal
angle $\phi$ between the transverse momenta of $\eta_{c},\gamma$
and the invariant mass $M_{\gamma\eta_{c}}$. In the Figure~\ref{fig:xiRapidity} we also have shown the relation
of the variables $\xi,\,\alpha_{\eta_{c}}$ to rapidities $y_{\eta_{c}},y_{\gamma}$.

In the generalized Bjorken kinematics, the variables $m_{N},\,\left|\Delta_{\perp}\right|,t$
are negligibly small, whereas all the other variables are parametrically
large, $\sim M_{\eta_{c}}$. In this kinematics it is possible to
simplify the light-cone decomposition~(\ref{eq:q}-\ref{eq:k}) and
obtain approximate relations of the Mandelstam variables with variables
$\alpha_{\eta_{c}},\boldsymbol{p}_{\perp}$ as 
\begin{equation}
-t'\approx\alpha_{\eta_{c}}M_{\gamma\eta_{c}}^{2}-M_{\eta_{c}}^{2},\qquad-u'\approx\left(1-\alpha_{\eta_{c}}\right)M_{\gamma\eta_{c}}^{2},\qquad\boldsymbol{p}_{\perp}^{2}=\alpha_{\eta_{c}}\left(1-\alpha_{\eta_{c}}\right)M_{\gamma\eta_{c}}^{2},\qquad M_{\gamma\eta_{c}}^{2}\approx2s\xi\label{eq:KinApprox}
\end{equation}
The physical constraint for real photon momenta
\begin{equation}
t'=\left(q-k\right)^{2}=-2q\cdot k=-2\,\left|\boldsymbol{q}\right|\left|\boldsymbol{k}\right|\left(1-\cos\theta_{\boldsymbol{q},\,\boldsymbol{k}}\right)\le0,
\end{equation}
 implies that the variable $\alpha_{\eta_{c}}$ is bound by $\alpha_{\eta_{c}}\ge M_{\eta_{c}}^{2}/M_{\gamma\eta_{c}}^{2}$,
so for the variable $u'$ we get 
\begin{equation}
-u'\le\left(-u'\right)_{{\rm max}}=M_{\gamma\eta_{c}}^{2}-M_{\eta_{c}}^{2}-t.
\end{equation}
We also may check that the pairwise invariant masses 
\begin{align*}
\left(P_{{\rm out}}+p_{\eta_{c}}\right)^{2} & \approx M_{\eta_{c}}^{2}+s\alpha_{\eta_{c}}\left(1-\xi\right)=t'+s\alpha_{\eta_{c}},\qquad\left(P_{{\rm out}}+k\right)^{2}\approx s\left(1-\alpha_{\eta_{c}}\right)\left(1-\xi\right),
\end{align*}
remain large, which shows that the produced $\eta_{c}$ and $\gamma$
are well-separated kinematically from the recoil proton. 

 In what follows we prefer to use as independent variables $t,t',M_{\gamma\eta_{c}}$
since, as we will see below, the cross-section decreases homogeneously
as a function of these variables. The photoproduction cross-section
in terms of these variables may be represented as
\begin{equation}
\frac{d\sigma_{\gamma p\to\eta_{c}\gamma p}}{dt\,dt'\,dM_{\gamma\eta_{c}}}\approx\frac{\left|\mathcal{A}_{\gamma p\to\eta_{c}\gamma p}\right|^{2}}{128\pi^{3}W^{4}M_{\gamma\eta_{c}}}\label{eq:Photo}
\end{equation}
where $\mathcal{A}_{\gamma p\to\eta_{c}\gamma p}$ is the amplitude
of the process. The electroproduction cross-section in the small-$Q$
kinematics gets the dominant contribution from events with single-photon
exchange between leptonic and hadronic parts, and may be represented
as~\cite{Weizsacker:1934,Williams:1935,Budnev:1975poe}
\begin{equation}
\frac{d\sigma_{ep\to eM_{1}M_{2}p}}{d\ln W^{2}dQ^{2}\,dt\,dt'\,dM_{\gamma\eta_{c}}}\approx\frac{\alpha_{{\rm em}}}{\pi\,Q^{2}}\,\left(1-y+\frac{y^{2}}{2}-(1-y)\frac{Q_{{\rm min}}^{2}}{Q^{2}}\right)\frac{d\sigma_{\gamma p\to M_{1}M_{2}p}}{dt\,dt'\,dM_{\gamma\eta_{c}}},\label{eq:LTSep}
\end{equation}
where $Q_{{\rm min}}^{2}=m_{e}^{2}y^{2}/\left(1-y\right)$, $m_{e}$
is the mass of the electron and $y$ is the fraction of the electron
energy which passes to the virtual photon (the so-called inelasticity);
it may be related to the invariant energy $\sqrt{s_{ep}}$ of the
electron-proton collision as 
\begin{equation}
y=\frac{W^{2}+Q^{2}-m_{N}^{2}}{s_{ep}-m_{N}^{2}}.
\end{equation}

\subsection{The amplitude of the $\eta_{c}\gamma$ pair production}

\label{subsec:Amplitudes}In the chosen kinematics it is possible
to evaluate the amplitude $\mathcal{A}_{\gamma p\to\eta_{c}\gamma p}$
in the collinear factorization framework, and express it in terms
of the GPDs of the target~\cite{Diehl:2000xz,Diehl:2003ny,Guidal:2013rya,Boer:2011fh,Burkert:2022hjz}.
As usual, we will disregard the mass of the proton $m_{N}$ and all
components of the momentum transfer $\Delta^{\mu}$. Furthermore,
we will assume that the invariant mass $M_{\gamma\eta_{c}}$ is large
enough to exclude feed-down contributions from radiative decays of
higher state charmonia. Our evaluation resembles previous evaluation
of the gluonic coefficient function for $\gamma\pi^{0}$~\cite{GPD2x3:8,GPD2x3:9,Duplancic:2024opc,Nabeebaccus:2023},
however due to use of heavy quark mass limit and NRQCD instead of
pion wave function, the final result is materially different and does
not allow direct comparison. In Appendix~\ref{sec:CoefFunction}
we provide a more detailed evaluation, which takes into account the
relative motion of heavy quarks in charmonia, and reduces to results
of~\cite{GPD2x3:8,GPD2x3:9,Duplancic:2024opc,Nabeebaccus:2023} in
hypothetical limit of vanishing quark mass.

\begin{figure}
\includegraphics[scale=0.4]{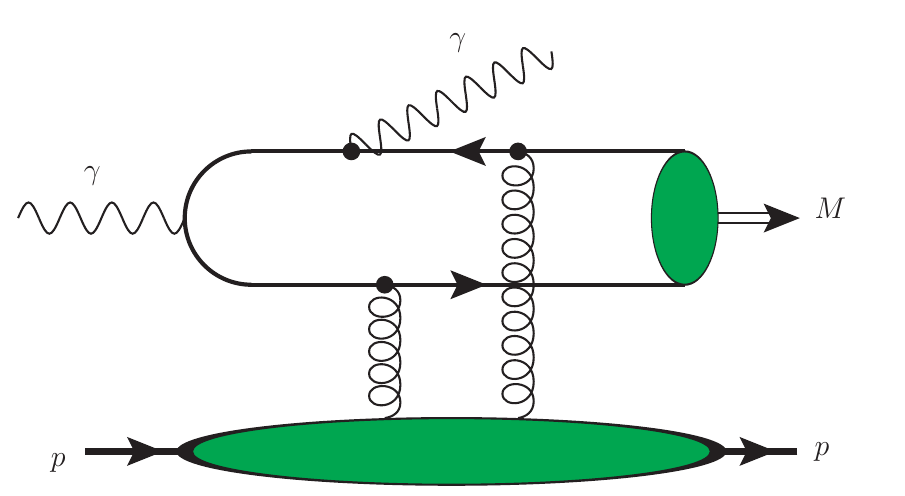}\includegraphics[scale=0.4]{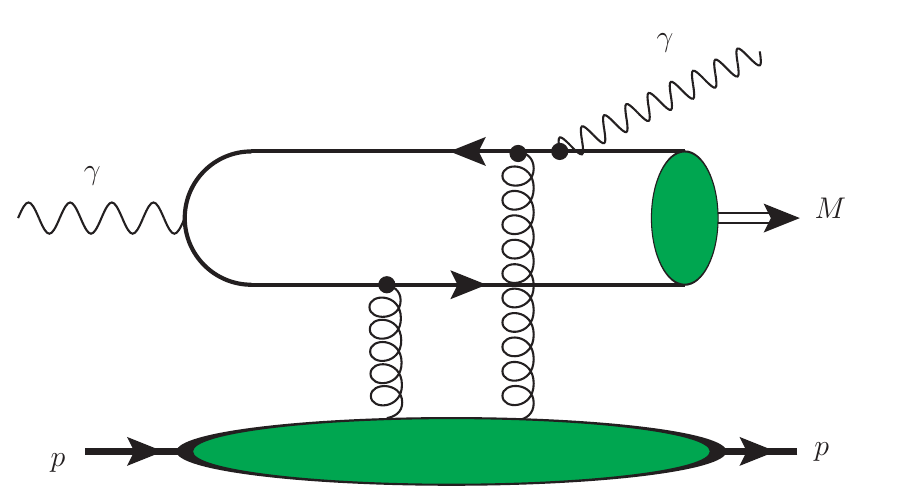}\includegraphics[scale=0.4]{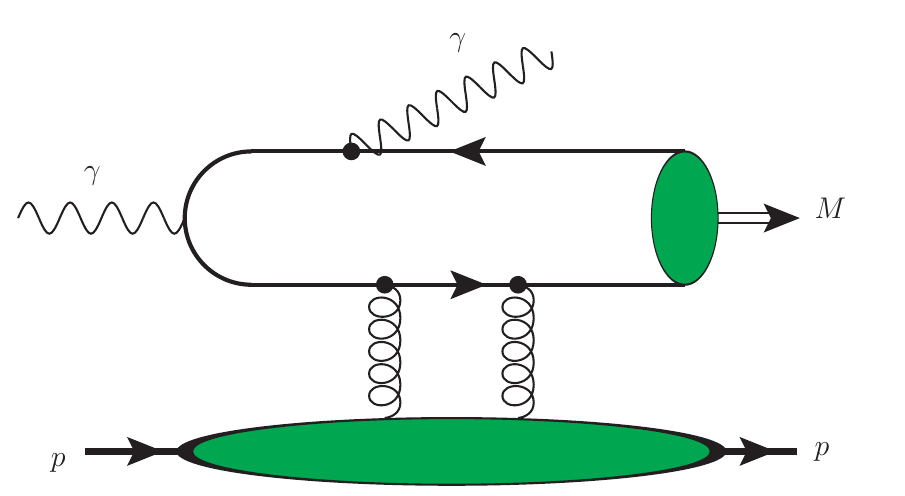}

\includegraphics[scale=0.4]{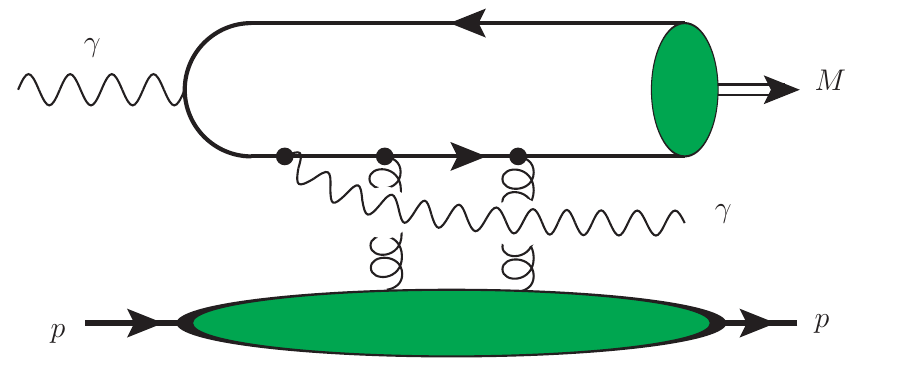}\includegraphics[scale=0.4]{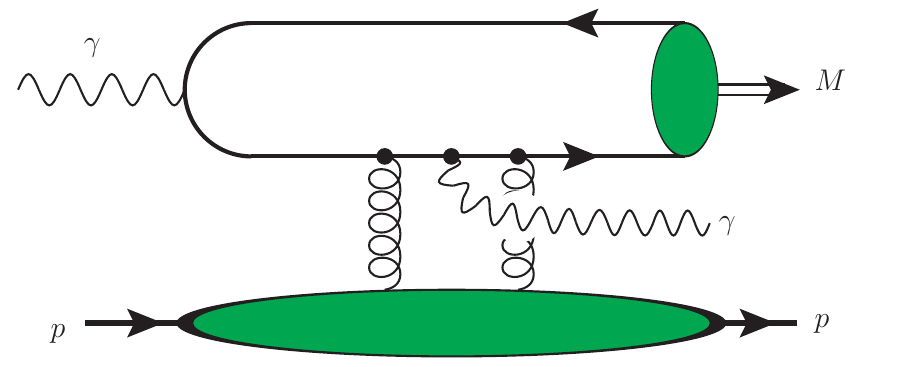}\includegraphics[scale=0.4]{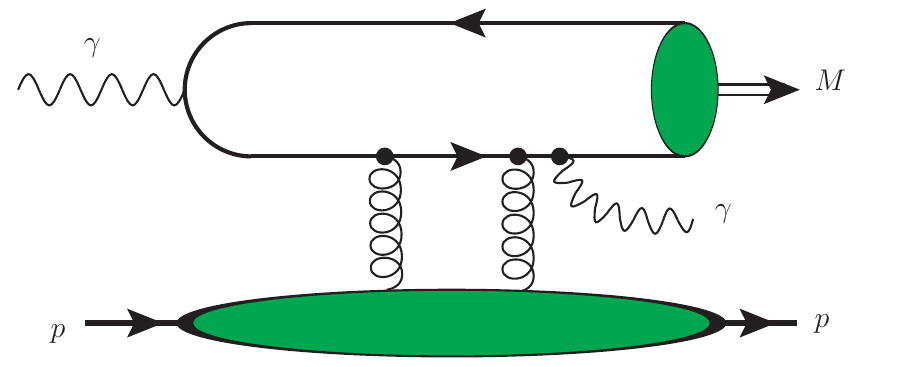}

\caption{\protect\label{fig:Photoproduction-A}The diagrams which contribute
to $\eta_{c}\gamma$ photoproduction. In all plots it is implied inclusion
of diagrams which may be obtained by inversion of heavy quark lines
(\textquotedblleft charge conjugation\textquotedblright ) and additional
contribution of the diagrams with permuted gluon vertices in $t$-channel
(24 diagrams in total, see Appendix~\ref{sec:CoefFunction} for more
details). }
\end{figure}

Since the GPDs are defined in the so-called symmetric frame, for evaluation
of the coefficient functions formally we should switch to that frame,
applying a transverse Lorentz boost 
\begin{align}
\ell^{+} & \to\ell^{+},\qquad\ell^{-}\to\ell^{-}+\ell^{+}\beta_{\perp}^{2}+\boldsymbol{\ell}_{\perp}\cdot\boldsymbol{\beta}_{\perp},\qquad\boldsymbol{\ell}_{\perp}\to\boldsymbol{\ell}_{\perp}+\ell^{+}\boldsymbol{\beta}_{\perp},\quad{\rm where}\quad\boldsymbol{\beta_{\perp}}=-\boldsymbol{\Delta_{\perp}}/\left(2\bar{P}_{{\rm SRF}}^{+}\right)
\end{align}
to the light-cone decomposition defined in~(\ref{eq:q}-\ref{eq:k}).
However, since the momentum $\boldsymbol{\Delta}_{\perp}$ eventually
can be disregarded when evaluating the coefficient function, we will
omit this step and will not distinguish the photon-proton~~(\ref{eq:q}-\ref{eq:k})
and the symmetric frames. The evaluation of the partonic-level amplitude
is straightforward. The momenta of the active parton (gluon) before
and after interaction are given explicitly by
\begin{align}
k_{i} & =\left((x+\xi)\bar{P}^{+},\,0,\,0\right),\quad k_{f}=\left((x-\xi)\bar{P}^{+},\,0,\,\boldsymbol{\Delta}_{\perp}\right)
\end{align}
where $x$ is the light-cone fraction of average momentum $\left(k_{i}^{+}+k_{f}^{+}\right)2\bar{P}^{+}$.
For unpolarized proton, the straightforward spinor algebra yields for
the square of the amplitude~\cite{Belitsky:2001ns}
\begin{align}
\sum_{{\rm spins}}\left|\mathcal{A}_{\gamma p\to M_{1}M_{2}p}\right|^{2} & =\left[4\left(1-\xi^{2}\right)\left(\mathcal{H}_{\gamma\eta_{c}}\mathcal{H}_{\gamma\eta_{c}}^{*}+\tilde{\mathcal{H}}_{\gamma\eta_{c}}\tilde{\mathcal{H}}_{\gamma\eta_{c}}^{*}\right)-\xi^{2}\left(\mathcal{H}_{\gamma\eta_{c}}\mathcal{E}_{\gamma\eta_{c}}^{*}+\mathcal{E}_{\gamma\eta_{c}}\mathcal{H}_{\gamma\eta_{c}}^{*}+\tilde{\mathcal{H}}_{\gamma\eta_{c}}\tilde{\mathcal{E}}_{\gamma\eta_{c}}^{*}+\tilde{\mathcal{E}}_{\gamma\eta_{c}}\tilde{\mathcal{H}}_{\gamma\eta_{c}}^{*}\right)\right.\label{eq:AmpSq}\\
 & \qquad\left.-\left(\xi^{2}+\frac{t}{4m_{N}^{2}}\right)\mathcal{E}_{\gamma\eta_{c}}\mathcal{E}_{\gamma\eta_{c}}^{*}-\xi^{2}\frac{t}{4m_{N}^{2}}\tilde{\mathcal{E}}_{\gamma\eta_{c}}\tilde{\mathcal{E}}_{\gamma\eta_{c}}^{*}\right],\nonumber 
\end{align}
where, inspired by previous studies of Compton scattering and single-meson
production~\cite{Belitsky:2001ns,Belitsky:2005qn}, we introduced
shorthand notations for the convolutions of the partonic amplitudes
with GPDs

\begin{align}
\mathcal{H}_{\gamma\eta_{c}}\left(\xi,\,t\right) & =\int dx\,C_{\gamma\eta_{c}}\left(x,\,\xi\right)H_{g}\left(x,\xi,t\right),\qquad\mathcal{E}_{\gamma\eta_{c}}\left(y_{1},y_{2},t\right)=\int dx\,C_{\gamma\eta_{c}}\left(x,\,\xi\right)E_{g}\left(x,\xi,t\right),\label{eq:Ha}\\
\tilde{\mathcal{H}}_{\gamma\eta_{c}}\left(\xi,\,t\right) & =\int dx\,\tilde{C}_{\gamma\eta_{c}}\left(x,\,\xi\right)\tilde{H}_{g}\left(x,\xi,t\right),\qquad\tilde{\mathcal{E}}_{\gamma\eta_{c}}\left(y_{1},y_{2},t\right)=\int dx\,\tilde{C}_{\gamma\eta_{c}}\left(x,\,\xi\right)\tilde{E}_{g}\left(x,\xi,t\right).\label{eq:ETildeA}
\end{align}
The partonic amplitudes $C_{\gamma\eta_{c}},\,\tilde{C}_{\gamma\eta_{c}}$
can be evaluated perturbatively, taking into account the diagrams
shown in the Figure~\ref{fig:Photoproduction-A}. In general the
coefficient functions depend on polarization vectors of the incoming
and outgoing photon. Since for the onshell photon there are only two
polarizations, we may conclude that for the $\gamma p\to\eta_{c}\gamma p$
there are only two independent helicity amplitudes, without and with
helicity flip. In what follows we will use additional superscript
notations $C_{\gamma\eta_{c}}^{(++)}$ and $C_{\gamma\eta_{c}}^{(+-)}$
to distinguish them. The final result for the coefficient functions
has a simple form (see Appendix~\ref{sec:CoefFunction} for details)

\begin{align}
C_{\gamma\eta_{c}}^{(++)} & \left(r,\,\alpha,\,\zeta=x/\xi\right)=\frac{4\bar{\alpha}\mathfrak{C}_{\eta_{c}}}{\xi^{2}\alpha^{2}\left(1+\bar{\alpha}\right)^{2}\left(r^{2}-1\right)\left(\bar{\alpha}+1/r^{2}\right)}\times\label{eq:Monome}\\
 & \times\frac{1}{\left(\zeta+1-i0\right)\left(\zeta-1+i0\right)\left(\zeta+\kappa-i0\right)\left(\zeta-\kappa+i0\right)}\left[\frac{}{}\text{\ensuremath{\alpha}}^{2}\left(1+\bar{\alpha}\right)\left(\zeta^{2}+1\right)+\right.\nonumber \\
 & +\frac{\left(\text{\ensuremath{\alpha}}^{2}-4\right)\zeta^{4}-2\left(\text{\ensuremath{\alpha}}^{3}-4\text{\ensuremath{\alpha}}+2\right)\zeta^{2}-\text{\ensuremath{\alpha}}\left(2\text{\ensuremath{\alpha}}^{2}+\text{\ensuremath{\alpha}}-4\right)}{\left(\zeta+1-i0\right)\left(\zeta-1+i0\right)r^{2}}+\frac{\left(1+\bar{\alpha}\right)^{2}\zeta^{4}+\text{\ensuremath{\alpha}}(\alpha^{2}+3\alpha-12)\zeta^{2}+\text{\ensuremath{\alpha}}(3\text{\ensuremath{\alpha}}^{2}+8\alpha-4)}{\left(\zeta+1-i0\right)\left(\zeta-1+i0\right)r^{4}}\nonumber \\
 & +\frac{-\text{\ensuremath{\alpha}}\left((\text{\ensuremath{\alpha}}-4)\zeta^{2}+11\text{\ensuremath{\alpha}}+8\right)+8\zeta^{2}+4}{\left(\zeta+1-i0\right)\left(\zeta-1+i0\right)r^{6}}-\frac{4\left(\zeta^{2}-3\text{\ensuremath{\alpha}}\right)}{\left(\zeta+1-i0\right)\left(\zeta-1+i0\right)r^{8}}\left.-\frac{4}{\left(\zeta+1-i0\right)\left(\zeta-1+i0\right)r^{10}}\right]\nonumber 
\end{align}
\begin{align}
C_{\gamma\eta_{c}}^{(+-)} & \left(r,\,\alpha,\,\zeta=x/\xi\right)=\frac{4\mathfrak{C}_{\eta_{c}}}{\xi^{2}\alpha^{2}\left(1+\bar{\alpha}\right)^{2}\left(r^{2}-1\right)\left(\bar{\alpha}+1/r^{2}\right)\left(\zeta+1-i0\right)^{2}\left(\zeta-1+i0\right)^{2}\left(\zeta+\kappa-i0\right)\left(\zeta-\kappa+i0\right)}\times\label{eq:Monome-2}\\
 & \times\left[\frac{\alpha^{2}\left(3\alpha^{2}+\alpha-3\right)-2\bar{\alpha}\left(\alpha^{3}-4\alpha+2\right)\zeta^{2}-\left(1+\bar{\alpha}\right)^{2}\left(1-\alpha\bar{\alpha}\right)\zeta^{4}}{r^{2}}+\right.\nonumber \\
 & -\frac{\alpha(3\alpha^{3}-2\alpha+5\alpha^{2}-8)-\bar{\alpha}\left(\alpha^{3}+3\alpha^{2}-8\right)\zeta^{2}+\bar{\alpha}\left(1+\bar{\alpha}\right)^{2}\zeta^{4}}{r^{4}}\nonumber \\
 & +\frac{-\bar{\alpha}\alpha(\alpha+4)\zeta^{2}+\alpha(\alpha+3)(11\alpha-4)-4}{r^{6}}-\frac{4\left(3\alpha^{2}+7\alpha-2-\bar{\alpha}\zeta^{2}\right)}{r^{8}}\left.+\frac{4(\alpha+3)}{r^{10}}\right]\nonumber 
\end{align}
\begin{align}
\tilde{C}_{\gamma\eta_{c}}^{(++)} & \left(r,\,\alpha,\,\zeta=x/\xi\right)=-\frac{4\zeta\bar{\alpha}\mathfrak{C}_{\eta_{c}}}{\xi^{2}\alpha^{2}\left(1+\bar{\alpha}\right)^{2}\left(r^{2}-1\right)\left(\bar{\alpha}+1/r^{2}\right)}\times\label{eq:Monome-1-1}\\
 & \frac{1}{\left(\zeta+1-i0\right)\left(\zeta-1+i0\right)\left(\zeta+\kappa-i0\right)\left(\zeta-\kappa+i0\right)}\left[\frac{}{}2\alpha\left(1+\bar{\alpha}^{2}\right)+\right.\nonumber \\
 & -\frac{2\left(\alpha^{3}-9\alpha^{2}+10\alpha-4+\left(2-5\bar{\alpha}-\bar{\alpha}^{3}\right)\zeta^{2}\right)}{\left(\zeta+1-i0\right)\left(\zeta-1+i0\right)r^{2}}-\frac{\left(3\alpha^{3}-17\alpha^{2}-8\alpha+16+\left(6-\bar{\alpha}-\bar{\alpha}^{3}\right)\zeta^{2}\right)}{\left(\zeta+1-i0\right)\left(\zeta-1+i0\right)r^{4}}\nonumber \\
 & -\frac{\bar{\alpha}\left(3\alpha^{2}-40\alpha+16+\left(1+\bar{\alpha}\right)^{2}\zeta^{2}\right)}{\left(\zeta+1-i0\right)\left(\zeta-1+i0\right)r^{6}}-\frac{2(\alpha+5)}{\left(\zeta+1-i0\right)\left(\zeta-1+i0\right)r^{8}}\left.+\frac{4}{\left(\zeta+1-i0\right)\left(\zeta-1+i0\right)r^{10}}\right]\nonumber 
\end{align}
\begin{align}
\tilde{C}_{\gamma\eta_{c}}^{(+-)} & \left(r,\,\alpha,\,\zeta=x/\xi\right)=\frac{4\zeta\mathfrak{C}_{\eta_{c}}}{\xi^{2}\alpha^{2}\left(1+\bar{\alpha}\right)^{2}\left(r^{2}-1\right)\left(\bar{\alpha}+1/r^{2}\right)\left(\zeta+1-i0\right)^{2}\left(\zeta-1+i0\right)^{2}\left(\zeta+\kappa-i0\right)\left(\zeta-\kappa+i0\right)}\times\label{eq:Monome-1}\\
 & \times\left[\frac{2\alpha\left(-2\alpha^{2}-\alpha+2+(2\alpha^{3}-8\alpha^{2}+11\alpha-4)\zeta^{2}\right)}{r^{2}}-\frac{\alpha\left(3\alpha^{3}-16\alpha^{2}-7\alpha+8\right)+\left(\alpha^{4}-9\alpha^{2}+16\alpha-4\right)\zeta^{2}}{r^{4}}+\right.\nonumber \\
 & +\frac{\alpha(3\alpha^{2}-35\alpha+8)-\bar{\alpha}\left(1+\bar{\alpha}\right)^{2}\zeta^{2}}{r^{6}}+\frac{4\left(\alpha^{2}+4\alpha-1\right)}{r^{8}}\left.+\frac{4\bar{\alpha}}{r^{10}}\right]\nonumber 
\end{align}
where we introduced shorthand notations $\alpha\equiv\alpha_{\eta_{c}}=(-u')/M_{\gamma\eta_{c}}$,
 $\zeta=x/\xi$, the constant $\mathfrak{C}_{\eta_{c}}$ in the prefactors
of~(\ref{eq:Monome}-\ref{eq:Monome-1}) is given explicitly by
\begin{equation}
\mathfrak{C}_{\eta_{c}}=\frac{4}{9N_{c}m_{c}}\pi^{2}\alpha_{{\rm em}}\alpha_{s}\left(\mu\right)\sqrt{\left\langle \mathcal{O}_{\eta_{c}}\right\rangle /m_{c}^{3}},\label{eq:CDef}
\end{equation}
$\left\langle \mathcal{O}_{\eta_{c}}\right\rangle \equiv\left\langle \mathcal{O}_{\eta_{c}}\left[^{1}S_{0}^{[1]}\right]\right\rangle \approx0.3\,{\rm GeV^{3}}$
is the color singlet long-distance matrix element (LDME) of $\eta_{c}$~\cite{Braaten:2002fi},
and $r$ is defined as
\begin{equation}
r=M_{\gamma\eta_{c}}/M_{\eta_{c}}=\frac{W}{M_{\eta_{c}}}\sqrt{\frac{2\xi}{1+\xi}\left(1-\frac{m_{N}^{2}}{W^{2}}\right)}\approx\frac{W}{M_{\eta_{c}}}\sqrt{\frac{2\xi}{1+\xi}}
\end{equation}
We also may observe that at large $r$ the photon helicity flip components
$C_{\gamma\eta_{c}}^{(+-)},\tilde{C}_{\gamma\eta_{c}}^{(+-)}$ have
a relative suppression by factor $1/r^{2}$ compared to non-flip components
$C_{\gamma\eta_{c}}^{(++)},\tilde{C}_{\gamma\eta_{c}}^{(+-)}$. This
implies a strong suppression of these components in the high energy
limit. The shorthand notation $\kappa$ which appears in the denominators
of the first lines of~(\ref{eq:Monome}-\ref{eq:Monome-1}) is defined
as 
\begin{equation}
\kappa=1-\frac{1-1/r^{2}}{1-\alpha/2}=\frac{1}{r^{2}}\frac{2-\alpha r^{2}}{2-\alpha}.\label{eq:kappa}
\end{equation}
It is possible to check that in physically relevant kinematics ($r>1,\,\alpha\in(0,1)$)
its values are limited by $|\kappa|<1$. The coefficient functions~(\ref{eq:Monome}-\ref{eq:Monome-1})
have poles in ERBL region at $\zeta=\pm1\mp i0$ and $\zeta=\pm\kappa\mp i0$
($x=\pm\xi\mp i0$ and $x=\pm\kappa\xi\mp i0$ in conventional notations).
The contour deformation prescription near these poles was found from
conventional $m^{2}\to m^{2}-i0$ prescription in Feynman propagators.
Some terms in~(\ref{eq:Monome}-\ref{eq:Monome-1}) include ordinary
second order poles $1/\left(x\mp\xi\pm i0\right)^{2}$, however they
do not present any difficulties for convergence of integral: formally,
the integration in the vicinity of these poles can be carried out
using an identity
\begin{align}
\int_{-1}^{1}dx\frac{F\left(x,\xi,t\right)}{\left(x-\xi+i0\right)^{2}} & =-\int_{-1}^{1}dx\,F\left(x,\xi,t\right)\frac{d}{dx}\frac{1}{\left(x-\xi+i0\right)}=-\left.\frac{F\left(x,\xi,t\right)}{\left(x-\xi+i0\right)}\right|_{x=-1}^{x=+1}+\int_{-1}^{1}dx\,\frac{\partial_{x}F\left(x,\xi,t\right)}{\left(x-\xi+i0\right)}.\label{eq:DoublePole}
\end{align}

\begin{figure}

\includegraphics[width=10cm]{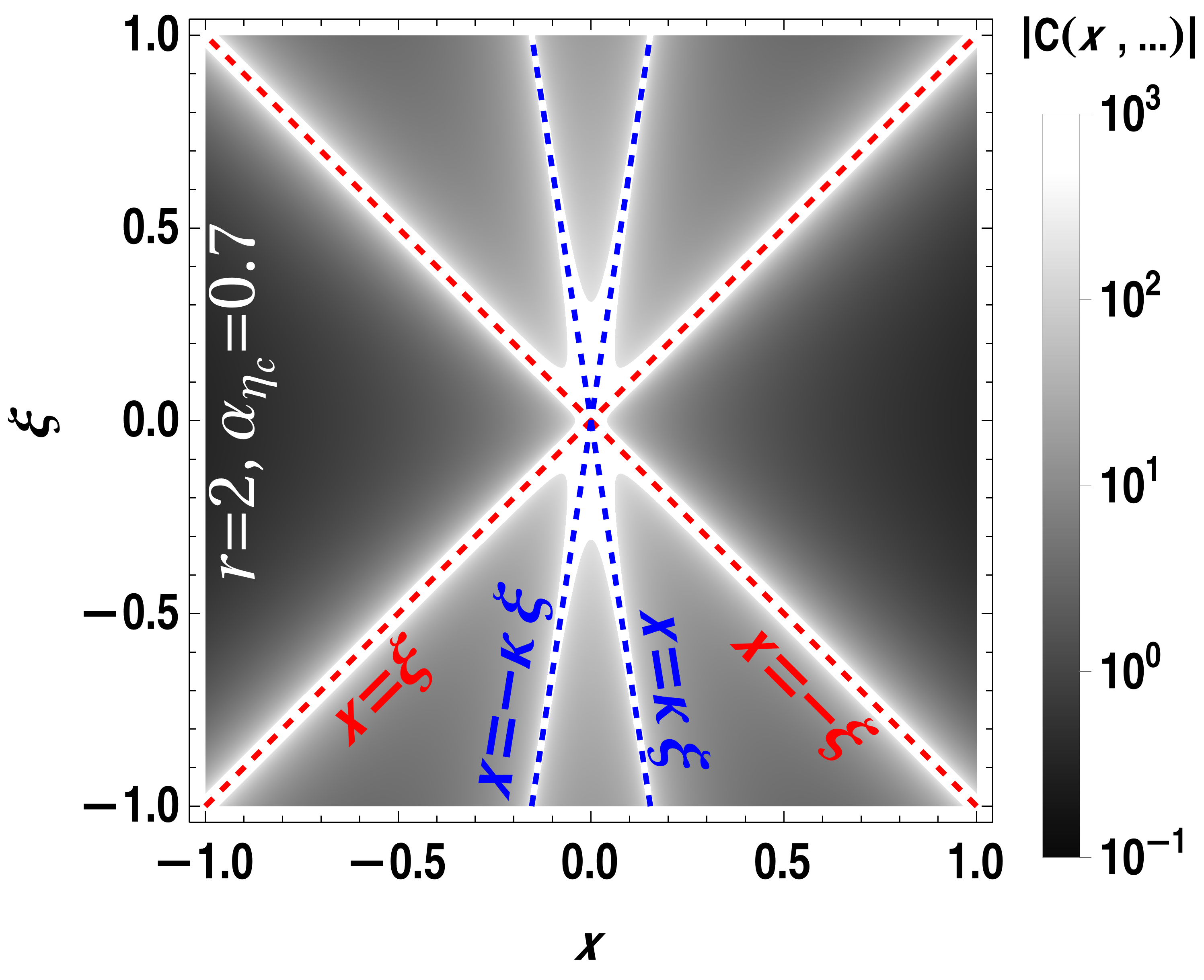}\includegraphics[width=8cm]{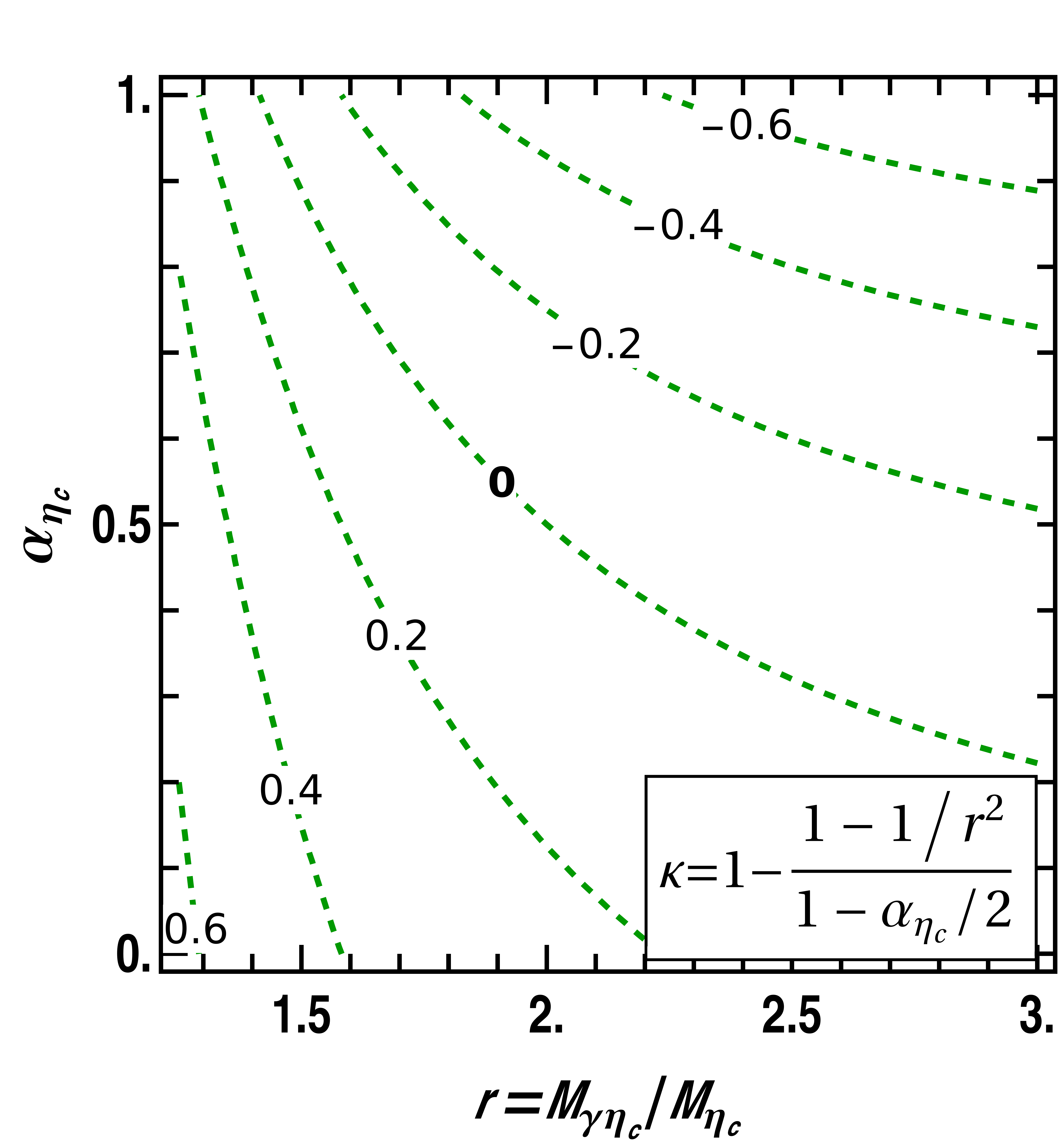}\caption{\protect\label{fig:CoefFunction} Left: Density plot which illustrates
the coefficient function $C_{\gamma\eta_{c}}^{(++)}$ (in relative
units) as a function of variables $x,\,\xi$. White lines effectively
demonstrate the position of the poles $x=\pm\xi,\,x=\pm\kappa\,\xi$
in the coefficient function~(\ref{eq:Monome}). The value of the
slope parameter $\kappa$ depends on kinematics and is defined in~(\ref{eq:kappa}).
Right: Contour plot which shows the dependence of the slope parameter
$\kappa$ on the choice of the kinematic variables $r=M_{\gamma\eta_{c}}/M_{\eta_{c}}$,
$\alpha_{\eta_{c}}=(-u')/M_{\gamma\eta_{c}}$. The parameter $\kappa$
is always is bound by a constraint $|\kappa|<1$ in the physically
relevant kinematics.}
\end{figure}

In the Figure~\ref{fig:CoefFunction} we show the density plot which
illustrates the behavior of  $C_{\gamma\eta_{c}}^{(++)}$ as a function
of its arguments. We can see that the function is strongly concentrated
in the vicinity of its poles, and decreases rapidly when we move away
from them. This implies that position of the poles determines the
region which gives the dominant contribution in convolution integrals~(\ref{eq:Ha}-\ref{eq:ETildeA}).
Varying the invariant mass $M_{\gamma\eta_{c}}$ and the value of
$\alpha_{\eta_{c}}$, it is possible to change the parameter $\kappa$
in the region $|\kappa|<1$ and in this way probe the behavior of
the gluon GPDs $H_{g}$ in the whole ERBL kinematics.

\section{Numerical estimates}

\label{sec:Numer}

\subsection{Differential cross-sections}

\label{subsec:diff}In what follows for the sake of definiteness,
we will use for our estimates the Kroll-Goloskokov parametrization
of the GPDs~\cite{Goloskokov:2006hr,Goloskokov:2007nt,Goloskokov:2008ib,Goloskokov:2009ia,Goloskokov:2011rd}.
As could be seen from~(\ref{eq:Photo},\ref{eq:AmpSq}), the cross-section
of the process includes contributions of GPDs with different helicity
states, however for unpolarized target the dominant contributions
stems from the GPD $H^{g}\left(x,\xi,t\right)$. 

\begin{figure}
\includegraphics[width=9cm]{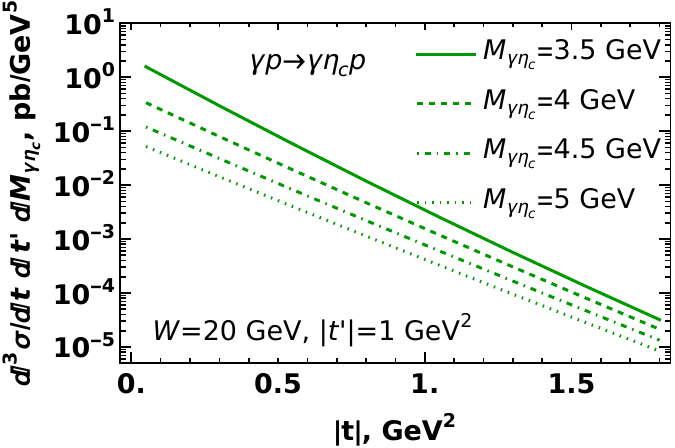}\includegraphics[width=9cm]{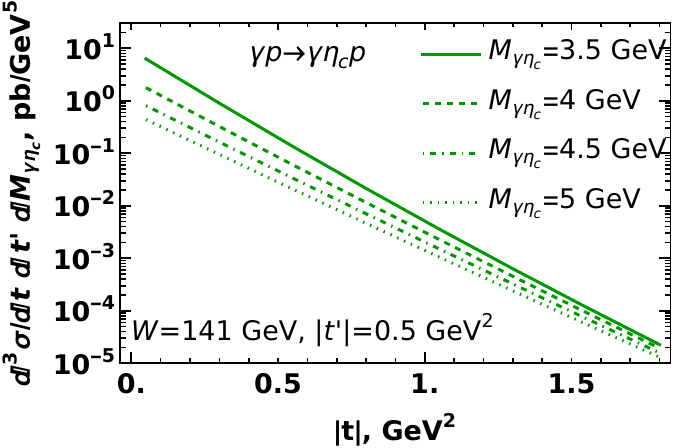}

\includegraphics[width=9cm]{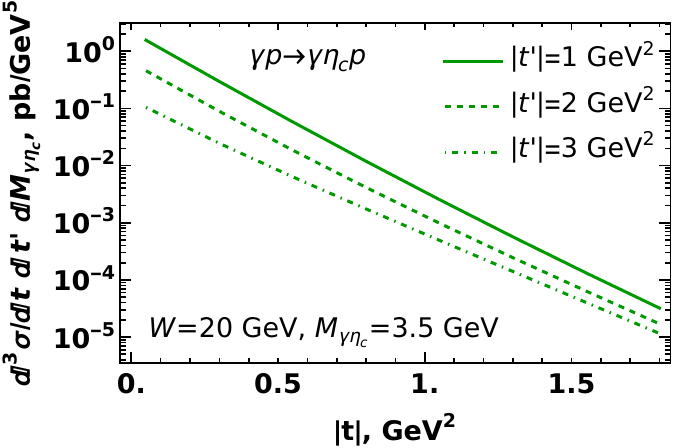}\includegraphics[width=9cm]{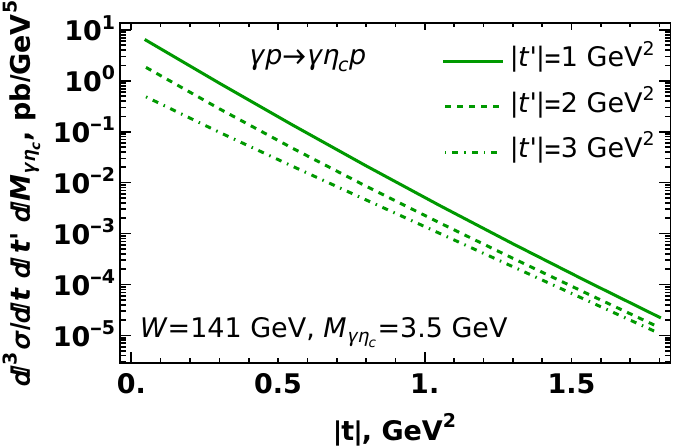}

\caption{\protect\label{fig:tDep}Dependence of the photoproduction cross-section~(\ref{eq:Photo})
on the invariant momentum transfer $t$ to the target at fixed invariant
mass $M_{\gamma\eta_{c}}$ (upper row) and fixed $|t'|$ (lower row).
The left and right columns differ by value of invariant energy $W$.
For other energies, the $t$-dependence has similar shape. }
\end{figure}

We would like to start presentation of the results with discussion
of the threefold differential cross-section~(\ref{eq:Photo}) and
its dependence on the invariant momentum transfer $t$ to the target.
In the collinear kinematics the variable $|t|$ is small and is disregarded
when evaluating the coefficient functions, for this reason the dependence
of the cross-section on this variable is largely due to the implemented
gluon GPD. In the Figure~\ref{fig:tDep} we show this dependence
for several invariant masses $M_{\gamma\eta_{c}}$ and choices of
the variable $t'$. A sharply decreasing $t$-dependence is common
to many phenomenological parametrizations of the GPDs and was introduced
to describe the pronounced $t$-dependence seen in DVCS and DVMP data.
For the $\eta_{c}\gamma$ photoproduction this implies that photon
and $\eta_{c}$ predominantly are produced with oppositely directed
transverse momenta $\boldsymbol{p}_{\eta_{c}}^{\perp},\,\boldsymbol{p}_{\gamma}^\perp$.
In view of a simple and well-understood dependence on $t$, in what
follows we will tacitly assume that $|t|=|t_{{\rm min}}|$, or consider
the observables in which the dependence on $t$ is integrated out.

\begin{figure}
\includegraphics[width=9cm]{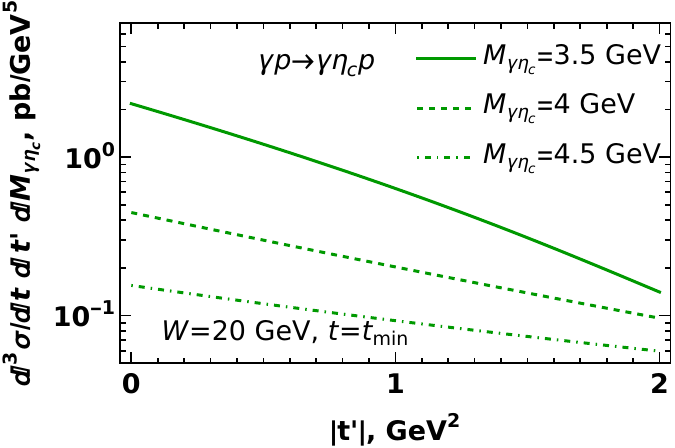}\includegraphics[width=9cm]{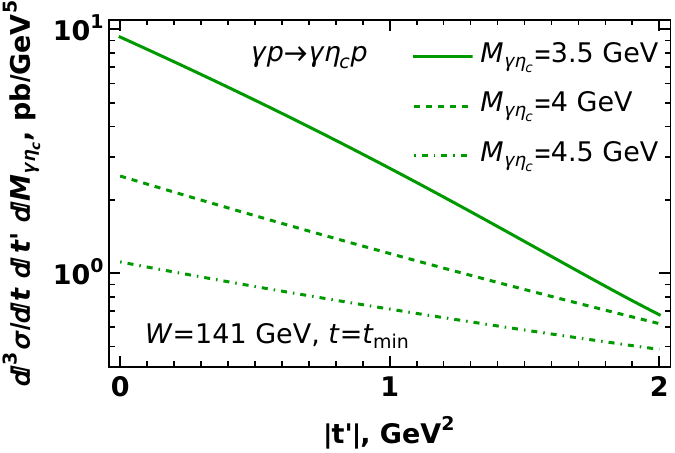}

\caption{\protect\label{fig:tPrimeDep}Dependence of the photoproduction cross-section~(\ref{eq:Photo})
on the variable $|t'|$ defined in~~(\ref{eq:Photo}). The left
and right plot correspond to different invariant energies $W=20$
GeV and $W=141$ GeV respectively.}
\end{figure}

In the Figure~\ref{fig:tPrimeDep} we show the dependence of the
cross-section~(\ref{eq:Photo}) on the variable $\left|t'\right|$.
Similar to the previous case, the cross-section decreases rapidly
as a function of this variable. As could be seen from~(\ref{eq:KinApprox}),
the dependence on $|t'|$ at fixed invariant mass $M_{\gamma\eta_{c}}^{2}$
is unambiguously related to $\alpha_{\eta_{c}}$-dependence of the
coefficient functions~(\ref{eq:Monome}-\ref{eq:Monome-1}). The
pefactor $\sim1/\alpha_{\eta_{c}}^{2}$ which appears in front of
all coefficient functions is partially responsible for the observed
$t'$-dependence of the cross-section. 

\begin{figure}
\includegraphics[width=9cm]{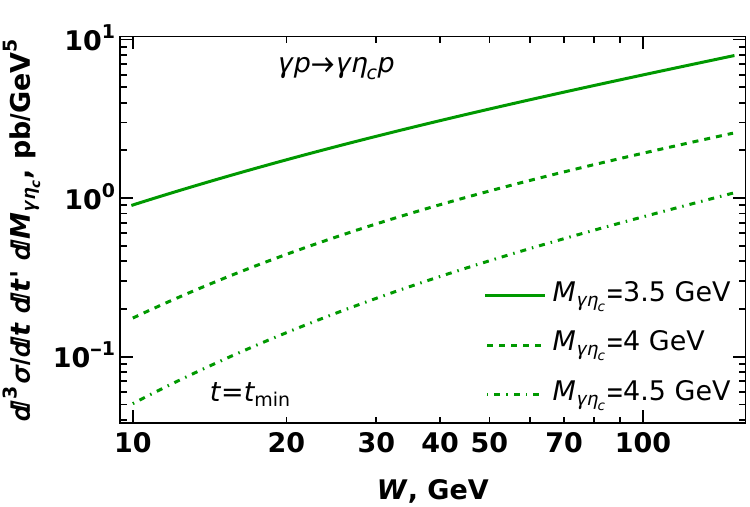}\includegraphics[width=9cm]{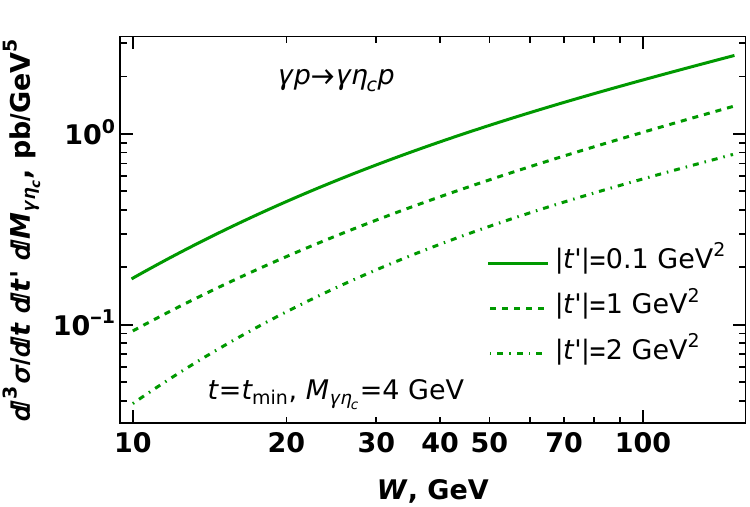}\caption{\protect\label{fig:WDep} Dependence of the cross-section on the invariant
collision energy $W$ at different values of the invariant mass $M_{\gamma\eta_{c}}$
(left) and variable $|t'|$ (right). The energy dependence of the
cross-section may be approximated as $d\sigma(W)/dt\,dt'\,dM_{\gamma\eta_{c}}\sim W^{\lambda}$,
where the slope parameter $\lambda$ has a very mild dependence on
other kinematic variables.}
\end{figure}

In the Figure~\ref{fig:WDep} we show the dependence of the cross-section
on the invariant energy $W$. As expected, the cross-section grows
with $W$, and nearly linear dependence in double logarithmic coordinates
suggests that the cross-section has a power-like dependence on energy,
$d\sigma(W)\sim W^{\lambda}$, where the parameter $\lambda$ has
a mild dependence on other kinematic variables. This behavior may
be understood if we take into account that the amplitude gets the
dominant contribution from the gluon GPD near $|x|\sim\xi\sim1/W^{2}$,
and the fact that the gluon GPDs grows as a function of the light-cone
fraction $x$. In the implemented phenomenological parametrization,
the small-$x$ behavior of the GPD roughly may be approximated as
$H_{g}(x,\xi,t)\sim x^{-\delta_{g}}$~\cite{Goloskokov:2006hr}.
After convolution with coefficient functions~(\ref{eq:Monome}-\ref{eq:Monome-1}),
this translates into a power-like behavior for the cross-section,
\begin{equation}
\frac{d\sigma(W)}{dt\,dt'\,dM_{\gamma\eta_{c}}}\sim\xi^{-2\delta_{g}}\sim W^{4\delta_{g}},\qquad\lambda=4\delta_{g},\label{eq:diffW}
\end{equation}
in agreement with results of numerical evaluation. The typical values
of the parameter $\lambda$ are $\lambda\approx0.7-0.8$.

\begin{figure}
\includegraphics[width=9cm]{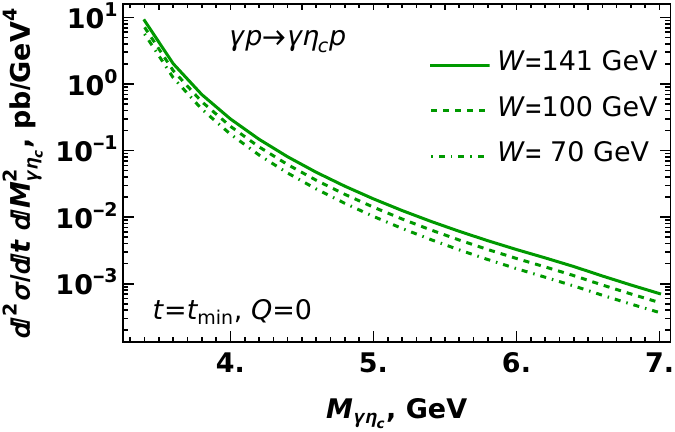}\includegraphics[width=9cm]{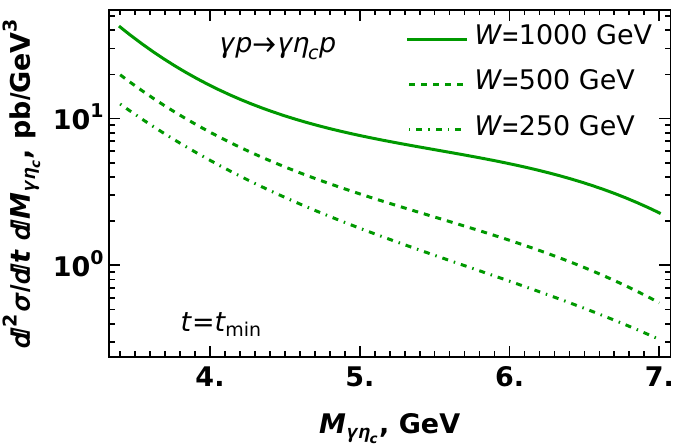}\caption{\protect\label{fig:M12Dep} Dependence of the cross-section on the
invariant masses of produced $\gamma\eta_{c}$ pair, for several proton
energies in the kinematics of EIC (left) and ultraperipheral collisions
at LHC (right). The right plot should be considered only as order
of magnitude estimate due to possible NLO corrections and saturation
effects. }
\end{figure}

In the Figure~\ref{fig:M12Dep} we show the dependence of the cross-section
on the invariant mass $M_{\gamma\eta_{c}}$. The dependence is very
pronounced and is a consequence of prefactor $\sim1/\left(r^{2}-1\right)=M_{\eta_{c}}^{2}/\left(M_{\gamma\eta_{c}}^{2}-M_{\eta_{c}}^{2}\right)$
in coefficient functions~~(\ref{eq:Monome}-\ref{eq:Monome-1})
and additional factor $\sim1/M_{\gamma \eta_c}$ in the cross-section~(\ref{eq:Photo}). Similar dependence on invariant mass
was also observed in other photon-meson production channels in the kinematics where invariant mass is large.

\begin{figure}
\includegraphics[width=9cm]{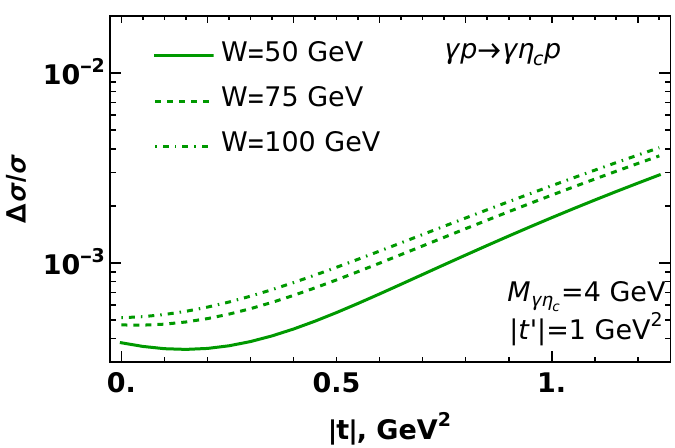}\includegraphics[width=9cm]{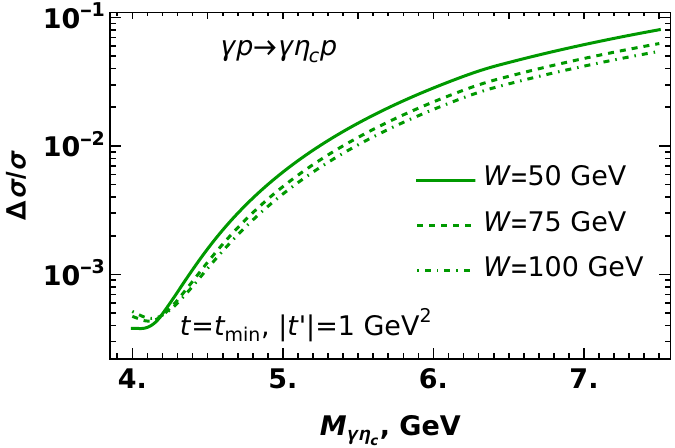}

\caption{\protect\label{fig:Polarized}The ratio of the cross-sections with
and without photon helicity flip, as a function of the momentum transfer
$|t|$ to the target and invariant mass of the $\eta_{c}\gamma$ pair.
The ratio remains small (below 10\%) in the whole range analyzed in
this paper.}
\end{figure}

Finally, we would like to discuss the relative contribution of different
polarizations of the final state photon to the total cross-section.
In the Figure~\ref{fig:Polarized} we plotted the ratio of the cross-sections
with and without the helicity flip of the final-state photon. As expected,
the ratio decreases as a function of energy $W$, though its dependence
on variables $|t|$ and $M_{\gamma\eta_{c}}$ is more complicated
and depends on the implemented parametrization of the GPD. However,
we can see that the ratio remains small (a few per cent or below)
in the region of small $|t|,\,M_{\gamma\eta_{c}}$ where the cross-section
is sufficiently large for experimental studies, and for this reason
predominantly the final-state and incoming photons will have the same
polarization. 

\subsection{Integrated cross-sections and counting rates}

\label{subsec:integr}
\begin{figure}
\includegraphics[totalheight=6cm]{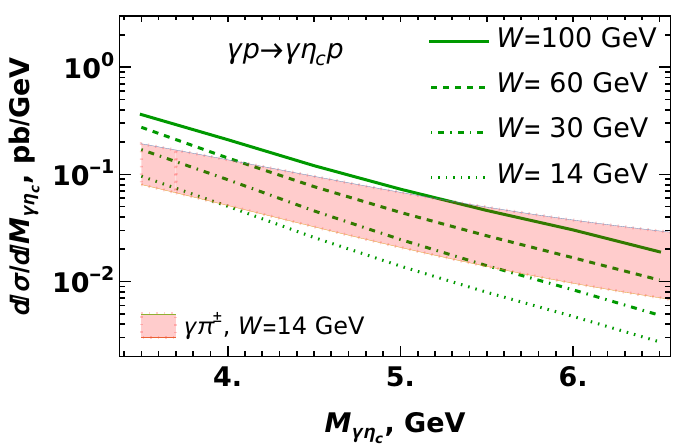}

\caption{\protect\label{fig:M12}The single-differential cross-section $d\sigma/dM_{\gamma\eta_{c}}$
as a function of the invariant mass of $\eta_{c}\gamma$ pair at different
collision energies. For reference we also added the colored band which
corresponds to the cross-section of $\gamma p\to\gamma\pi^{+}n$ at
$W=14$~GeV found in~\cite{GPD2x3:9}. }
\end{figure}

So far we considered the differential cross-sections,
which are best suited for theoretical studies. Unfortunately, it is
difficult to measure such small cross-sections because of insufficient
statistics, and for this reason now we will provide predictions for
the yields integrated over some or all kinematic variables. In the
left panel of the Figure~\ref{fig:M12} we have shown the cross-section
$d\sigma/dM_{\gamma\eta_{c}}$ for different energies. For the sake
of comparison we also added a colored band which corresponds to cross-section
of $\gamma p\to\gamma\pi^{+}n$ at $W=14$~GeV found in~\cite{GPD2x3:9}~\footnote{The predictions for $\gamma p\to\gamma\pi^{+}n$ were directly extracted
from Figure 16 in~\cite{GPD2x3:9} and converted to $d\sigma/dM_{\gamma M}$
units.}. The width of the band represents uncertainty due to choice of pion
distribution amplitudes and quark GPDs. We may see that the cross-sections
of $\eta_{c}\gamma$ and $\pi^{-}\gamma$ photoproduction have similar
magnitudes if compared at the same (large) values of the invariant
mass of photon-meson pair. 

\begin{figure}
\includegraphics[totalheight=6cm]{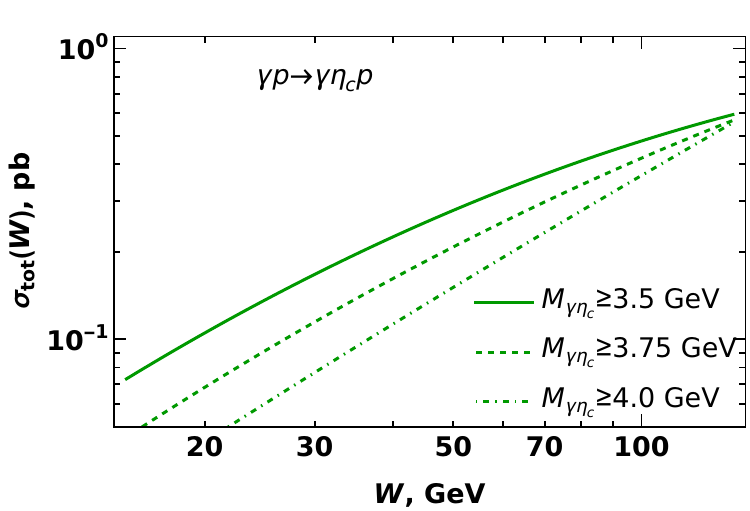}\includegraphics[totalheight=6cm]{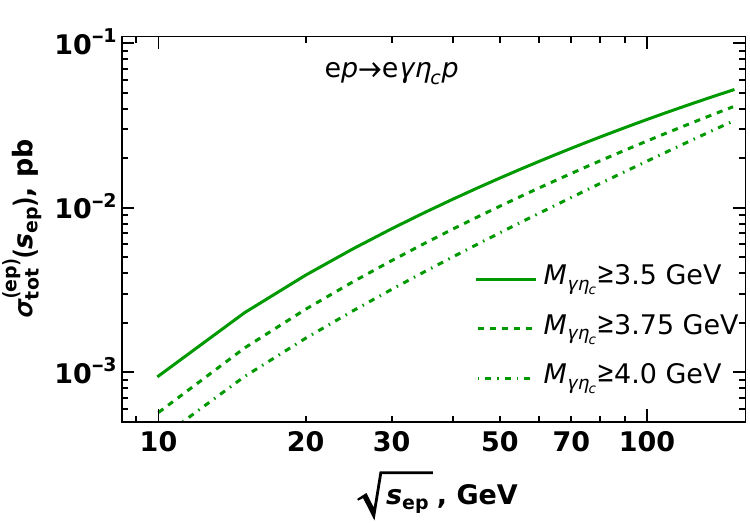}

\caption{\protect\label{fig:Total}Predictions for the total (\textquotedblleft fiducial\textquotedblright )
cross-section of photoproduction (left) and electroproduction (right)
of $\eta_{c}\gamma$ as a function of energy $W$ for different cutoffs
on the invariant mass of $\eta_{c}\gamma$ pairs.}
\end{figure}

In the left panel of the Figure~\ref{fig:Total} we show the energy
dependence of the total (fully integrated) cross-section $\sigma_{{\rm tot}}(W)$
for the photoproduction of $\eta_{c}\gamma$. As we discussed earlier,
the suggested mechanism may be applied only in the kinematics where
the invariant mass $M_{\gamma\eta_{c}}$ of the photon-meson pair
is sufficiently large in order to avoid the feed-down contributions
from radiative decays of the excited quarkonia states. The choice
of the minimal value $\left(M_{\gamma\eta_{c}}\right)_{{\rm min}}$
is somewhat arbitrary, for this reason we have shown the result for
several possible cutoffs. This dependence is stronger at small energies
due to phase space contraction for production of $\eta_{c}\gamma$
pair with large invariant mass. At large $W$, for all cutoffs the
$W$-dependence may be approximated as 
\begin{equation}
\sigma_{{\rm tot}}\left(W,\,M_{\gamma\eta_{c}}\ge3.5\,{\rm GeV}\right)\approx0.48\,{\rm pb}\,\left(\frac{W}{100\,{\rm GeV}}\right)^{0.75},
\end{equation}
in agreement with our earlier findings~(\ref{eq:diffW}) for the
energy dependence of the differential cross-sections. In the right
panel of the same Figure~\ref{fig:Total} we have shown our estimates
for the fully integrated cross-section of the electroproduction $ep\to e\gamma\eta_{c}p$ as a function
of the invariant electron-proton collision energy $\sqrt{s_{ep}}$.
For the upper EIC energy $\sqrt{s_{ep}}\approx141\,$GeV the corresponding
cross-section is 
\begin{equation}
\sigma_{{\rm tot}}^{({\rm ep})}\left(\sqrt{s_{ep}}=141\,{\rm GeV},\,M_{\gamma\eta_{c}}\ge3.5\,{\rm GeV}\right)\approx49\,{\rm fb}.\label{eq:XSecEIC}
\end{equation}
This smallness is due to $\sim\alpha_{{\rm em}}$ in the leptonic
prefactor (see~(\ref{eq:LTSep})) and a very steep slope of the $t$-dependence
in the exclusive process. For the instantaneous luminosity $\mathcal{L}=10^{34}\,{\rm cm^{-2}s^{-1}}=10^{-5}{\rm fb}^{-1}s^{-1}$
at the future Electron Ion Collider~\cite{Accardi:2012qut,AbdulKhalek:2021gbh,Navas:2024X}
the cross-section~(\ref{eq:XSecEIC}) gives a production rate $dN/dt\approx$42
events/day, with approximately $N=4.9\times10^{3}$ produced $\eta_{c}\gamma$
pairs per each $\int dt\,\mathcal{L}=100{\rm fb^{-1}}$ of integrated
luminosity. Since $\eta_{c}$ mesons are not detected directly but
rather via their decays into light hadrons, for analysis of feasibility
it is also interesting to know the counting rate $dN_{d}/dt$ and
the total number of detected events $N_{d}$ for a chosen decay mode.
Technically, these quantities may be found multiplying $dN/dt$ and
$N$ by the branching fraction of $\eta_{c}$ to the chosen decay
mode. In experimental studies the $\eta_{c}$-meson is frequently
identified via its decays to pions and kaons, e.g.: $\eta_{c}(1S)\to K_{S}^{0}K^{+}\pi^{-}$,
for which the corresponding branching fraction is~\cite{BESIII:2019eyx,Navas:2024X}
\begin{equation}
{\rm Br}_{\eta_{c}}={\rm Br}\left(\eta_{c}(1S)\to K_{S}^{0}K^{+}\pi^{-}\right)=2.6\%.
\end{equation}
This translates into detection (counting) rate $dN_{d}/dt\approx$32
events/month, with $N_{d}=$127 detected events per $100{\rm fb^{-1}}$
of integrated luminosity. 

For the kinematics of the ultraperipheral collisions at LHC, the counting
rates could be significantly larger due to increase of the cross-section
with energy. Furthermore, in $pA$ collisions the counting rate could
be additionally enhanced by factor $\sim Z^{2}$ in the photon flux,
where $Z$ is the atomic number of the projectile nucleus. However,
the suggested approach might be not applicable in that kinematics
due to onset of saturation effects, so we abstain from making detailed
predictions for LHC energies.

\subsection{Comparison with $\eta_{c}$ photoproduction}

\label{subsec:odderon}The exclusive $\eta_{c}$ photoproduction process
$\gamma p\to\eta_{c}p$ for a long time has been considered as one
of the most promising channels for studies of odderons~\cite{Odd4,Odd5,Odd6,Odd7}:
the $C$-odd 3-gluon exchanges in $t$-channel predicted in~\cite{Odd1,Odd2,Odd3}.
While the existence of the odderons has never been questioned, for
a long time the magnitude of the odderon-mediated processes remained
largely unknown because it is controlled by a completely new nonperturbative
amplitude (see~\cite{Odd7,Benic:2023} for a short overview). Only
recently the experimental measurements could confirm nonzero contribution
of odderons from comparison of $pp$ and $p\bar{p}$ elastic cross-sections
measured at LHC and Tevatron~\cite{OddTotem1,OddTotem2}. Since that
analysis potentially could include sizable uncertainties due to details
of extrapolation procedure, the searches of odderons shifted towards
channels which require the $C$-odd $t$-channel exchanges. The photoproduction
$\gamma p\to\eta_{c}p$ is a rather clean channel for study of odderons
using perturbative methods, and for this reason it will remain in
focus of future experimental studies, both at HL-LHC and at future
EIC.

The $\eta_{c}\gamma$ photoproduction in this context deserves a lot
of interest because it could constitute a sizable background to $\eta_{c}$
photoproduction. Indeed, the $\eta_{c}\gamma$ photoproduction does
not require small $C$-odd exhanges in $t$-channel, and the latter
fact potentially could compensate the expected $\mathcal{O}\left(\alpha_{{\rm em}}\right)$-suppression
of the cross-section. Since the acceptance of modern detectors for
photons is below unity, potentially the $\eta_{c}\gamma$ photoproduction
with undetected final-state photons could be misinterpreted as $\gamma p\to\eta_{c}p$
subprocess. The accurate estimate of the background in general requires
detailed knowledge of the detector's geometry and acceptance. For
the sake of simplicity we will assume that \emph{all} photons are
undetected, and will discuss the cross-section $d\sigma/dt$, integrating
over the phase space of the produced photon. Such approach provides
an upper estimate for the background.

\begin{figure}
\includegraphics[width=9cm]{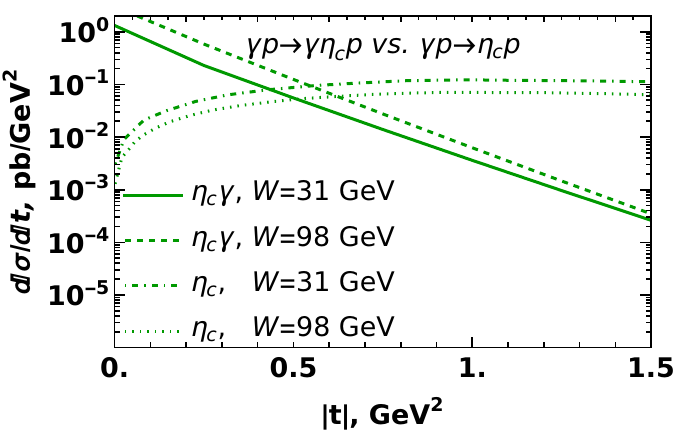}\caption{\protect\label{fig:Odd} Comparison of the cross-sections of the $\gamma p\to\eta_{c}p$
and $\gamma p\to\eta_{c}\gamma p$ with undetected (integrated out)
final-state photon. The cross-section of $\gamma p\to\eta_{c}p$ is
taken from~\cite{Odd7,Benic:2023} (curves $x=10^{-2}$ and $x=10^{-3}$,
the values of $W$ are restored assuming the relation $x=M_{\eta_{c}}^{2}/W^{2}$).}
\end{figure}

In the Figure~(\ref{fig:Odd}) we compare the cross-sections of the
$\eta_{c}$ and $\eta_{c}\gamma$ with undetected (integrated out)
photon in the final state. We can see that in the small-$t$ kinematics
($|t|\lesssim0.5\,{\rm GeV}^{2}$) the cross-section of $\eta_{c}\gamma$
exceeds the cross-section of the $\eta_{c}$ photoproduction, though
becomes suppressed exponentially at higher $t$ due to implemented
GPD model~\cite{Goloskokov:2006hr,Goloskokov:2007nt,Goloskokov:2008ib,Goloskokov:2009ia,Goloskokov:2011rd}.
Furthermore, the cross-section of $\eta_{c}\gamma$ grows faster with
energy and thus eventually will exceed the $\eta_{c}$ cross-section
at any $t$. In collinear factorization approach the energy dependence
is controlled by $x$-dependence of the GPD parametrization. In BFKL
language, the difference of energy dependencies of $\eta_{c}\gamma$
and $\eta_{c}$ is a consequence of the fact that pomeron has larger
intercept than odderon.

The fact that the odderon contribution in the small-$t$ kinematics
can be shadowed by other mechanisms has been discussed earlier in~\cite{Benic:2023}.
However, the corrections discussed in that paper depend on the isospin
of the target, and are significantly less important for processes
on neutrons, namely for $\gamma n\to\eta_{c}n$ subprocess. In contrast,
the cross-section of $\eta_{c}\gamma$ does not depend on isospin
of the target and will be the same for $\gamma n\to\eta_{c}\gamma n$
and $\gamma p\to\eta_{c}\gamma p$ subprocesses.

\section{Conclusions}

\label{sec:Conclusions}In this paper we analyzed the exclusive $\eta_{c}\gamma$
photoproduction in the collinear factorization approach. Our evaluation
in the massless limit agrees with earlier evaluations of the gluonic
coefficient function for $\gamma\pi^{0}$ photoproduction~\cite{GPD2x3:8,GPD2x3:9}.
The nonzero quark mass allows to avoid factorization breaking pinch
singularities which were discovered in~\cite{Duplancic:2024opc,Nabeebaccus:2023}
in massless case. The coefficient function of the process is given
by a simple meromorphic function of its arguments, with two ``classical''
poles at $x=\pm\xi$, and two additional poles at $x=\pm\kappa\,\xi$,
where the value of $\kappa$ depends on kinematics (values of $M_{\gamma\eta_{c}}$
and $u'$) and is bound by $|\kappa|<1$ in the physically permitted
kinematical range. The analytic structure of the coefficient function
apparently does not allow deconvolution (direct extraction of GPDs
from amplitudes). However, we observed that the coefficient function
is strongly peaked in the vicinity of its poles, and falls off rapidly
outside, so the cross-section of the process is largely determined
by the behavior of the gluon GPDs near these poles. Varying the kinematics
and in this way changing position of the poles, the suggested process
may be used to constrain the existing phenomenological parametrizations
of GPDs in the whole ERBL region. 

Numerically, the amplitude of the process obtains the dominant contribution
from the unpolarized gluon GPD $H_{g}$ in ERBL kinematics. The cross-section
of $\eta_{c}\gamma$ photoproduction is comparable to cross-sections
of the process $\gamma^{*}p\to\gamma Mp$, $M=\pi,\,\rho$ if we compare
them at the same (large) values of invariant masses of meson-photon
pair. However, due to inherently large value of the invariant mass
of the $\gamma\eta_{c}$ pair, the total (integrated) photoproduction
cross-section is below a picobarn. Nevertheless, we expect that this
cross-section is within the reach of experimental studies: for example,
for $ep$ collisions at EIC we estimate the production rate of a few
thousand $\eta_{c}\gamma$ pairs per each 100 ${\rm fb}^{-1}$ of
integrated luminosity. We observed that the cross-section grows with
energy and could reach large values for TeV-range photon-proton collisions,
however in that kinematics the suggested approach is not very reliable
due to expected large NLO corrections and onset of saturation effects.

 We expect that the suggested process could be studied in ultraperipheral
collisions at LHC, as well as in electron-proton collisions at the
future Electron Ion Collider (EIC)~\cite{Accardi:2012qut,AbdulKhalek:2021gbh,Burkert:2022hjz}
and possibly at JLAB after future 22 GeV upgrade~\cite{Accardi:2023chb}. 

\section*{Acknowldgements}

We thank our colleagues at UTFSM university for encouraging discussions.
We also thank M. Sanhueza for providing links to references~\cite{Odd7,Benic:2023}.
This research was partially supported by ANID PIA/APOYO AFB230003
(Chile) and Fondecyt (Chile) grant 1220242. \textquotedbl Powered@NLHPC:
This research was partially supported by the supercomputing infrastructure
of the NLHPC (ECM-02)\textquotedbl .

\appendix

\section{Evaluation of the coefficient functions}

\label{sec:CoefFunction}The evaluation of the coefficient functions
can be performed using the standard light--cone rules from~\cite{Lepage:1980fj,Brodsky:1997de,Diehl:2000xz,Diehl:2003ny,Diehl:1999cg,Ji:1998pc}.
As discussed in Section~\ref{subsec:Kinematics}, we disregard the
proton mass $m_{N}$ and the invariant momentum transfer to the proton
$t=\Delta^{2}$, treating all the other variables as parametrically
large quantities of order $m_{c}$. The evaluation of the partonic
amplitudes shown in the Figure~\ref{fig:Photoproduction-A} is straightforward. 

In order to compare our results with earlier papers on $\gamma\pi^{0}$
production, we will assume that the heavy quarks carry the fractions
$z$ and $1-z$ of the quark-antiquark momenta, though eventually
we will disregard the internal motion of the quarks inside $\eta_{c}$
and take the limit $z=1/2$. Formally, this can be done introducing
the light-cone distribution amplitude (LCDA), and relating it to NRQCD
LDMEs, as discussed in~\cite{Ma:2006hc,Wang:2013ywc,Wang:2017bgv}.
The definitions of quarkonia distribution amplitude is a straightforward
extensions of general results formulated for light mesons~\cite{LightQuarkDA:1,LightQuarkDA:2,LightQuarkDA:3,LightQuarkDA:4,LightQuarkDA:5}.
For the spinless $\eta_{c}$-meson, at leading twist there is only
one distribution amplitude defined as 
\begin{align}
\Phi_{\eta_{c}}\left(z\right) & =\int\frac{d\lambda}{2\pi}e^{izp^{+}\lambda}\left\langle 0\left|\bar{\psi}\left(-\frac{\lambda}{2}\right)\gamma^{+}\gamma_{5}\mathcal{L}\left(-\frac{\lambda}{2},\,\frac{\lambda}{2}\right)\psi\left(\frac{\lambda}{2}\right)\right|\eta_{c}(p)\right\rangle ,\label{eq:DAdefEta}\\
 & \mathcal{L}\left(-\frac{\lambda}{2},\,\frac{\lambda}{2}\right)\equiv\mathcal{P}{\rm exp}\left(i\int_{-\lambda/2}^{\lambda/2}d\zeta\,A^{+}\left(\zeta\right)\right).
\end{align}
where we assume that $\eta_{c}$ moves in the plus-direction, $z$
is the fraction of its momentum carried by the $c$-quark, $\mathcal{L}$
is the usual Wilson link. Sometimes the definition~(\ref{eq:DAdefEta})
is written for the distribution amplitudes normalized to unity and
thus may include additional normalization factor (decay constant $f_{\eta_{c}}$).
The NRQCD approach constructs the description of quarkonia states
in terms of long-distance matrix elements with structure $\left\langle 0\left|\hat{\mathcal{O}}\right|\eta_{c}(p)\right\rangle ,$
where $\hat{\mathcal{O}}$ is a set of \emph{local} operators built
from operators of quarks, antiquarks and their covariant derivatives.
The relation of the two approaches has been discussed in detail in~\cite{Ma:2006hc,Wang:2013ywc,Wang:2017bgv}.
The distribution amplitudes may be expressed in terms of NRQCD LDMEs
as 
\begin{equation}
\Phi_{\eta_{c}}(z)=\hat{\Phi}_{\eta_{c}}\left(z\right)\frac{\left\langle 0\left|\hat{\mathcal{O}}_{\eta_{c}}\right|\eta_{c}(p)\right\rangle }{2\sqrt{m_{c}}}\left(1+\mathcal{O}\left(v^{2}\right)\right),\label{eq:F}
\end{equation}
where $\hat{\Phi}\left(z\right)$ is the perturbative (partonic-level)
distribution amplitude, and the denominator is conventionally added
to take into account the difference between normalizations of Fock
states used in LCDA and NRQCD pictures. In the leading order over
relative velocity $v\sim\mathcal{O}\left(\alpha_{s}\left(m_{c}\right)\right)$
the distribution amplitude may be written as 
\begin{equation}
\hat{\Phi}_{\eta_{c}}\left(z\right)\sim f_{\eta_{c}}\delta\left(z-\frac{1}{2}\right)+\mathcal{O}\left(v^{2}\right).\label{eq:PhiHat}
\end{equation}
As was discussed in~\cite{Ma:2006hc}, the next-to-leading order
corrections to the amplitude~(\ref{eq:PhiHat}) give nontrivial dependence
on~$z$, however we will eventually disregard it, since formally
it is a higher-order correction in $\alpha_{s}$. Conversely, the
color singlet LDMEs~$\left\langle 0\left|\hat{\mathcal{O}}_{\eta_{c}}\right|\eta_{c}(p)\right\rangle $
may be expressed in terms of the moments of the distribution amplitude
$\Phi_{\eta_{c}}(z)$. 

In general, the coefficient functions $C_{\mathfrak{\gamma\eta_{c}}},\,\tilde{C}_{\mathfrak{\gamma\eta_{c}}}$
which appear in~(\ref{eq:Ha}-\ref{eq:ETildeA}) should be understood
as convolutions of partonic-level amplitudes with distribution amplitudes
$\Phi_{\eta_{c}}$, namely, 
\begin{align}
C_{\mathfrak{\gamma\eta_{c}}}\left(x,\,\xi\right) & =\int_{0}^{1}\,dz\,c\left(x,\,\xi,\,z\right)\Phi_{\eta_{c}}\left(z\right),\qquad\tilde{C}_{\gamma\eta_{c}}\left(x,\,\xi\right)=\int_{0}^{1}\,dz\,\tilde{c}\left(x,\,\xi,\,z\right)\Phi_{\eta_{c}}\left(z\right).\label{eq:ca}
\end{align}
The evaluation of the partonic amplitudes $c\left(x,\,\xi,\,z\right),\,\tilde{c}\left(x,\,\xi,\,z\right)$
may be done perturbatively. The projections onto a state with total
spin $S=0$ in the NRQCD picture may be found using proper Clebsch-Gordan
coefficients~\cite{Cho:1995ce,Cho:1995vh,DVMPcc1}, 
\begin{align}
\hat{P}_{SS_{z}}^{ij} & =-\sqrt{\frac{\left\langle \mathcal{O}_{\eta_{c}}\right\rangle }{m_{Q}}}\,\frac{1}{8m_{Q}}\left(\frac{\hat{p}_{\eta_{c}}}{2}-\frac{\hat{\delta}_{\eta_{c}}}{2}-m_{c}\right)\gamma_{5}\left(\frac{\hat{p}_{\eta_{c}}}{2}+\frac{\hat{\delta}_{\eta_{c}}}{2}+m_{c}\right)\otimes\frac{\delta_{ij}}{\sqrt{N_{c}}},\label{eq:NRQCDProjector}
\end{align}
where $\left\langle \mathcal{O}_{\eta_{c}}\right\rangle \equiv\left\langle \mathcal{\mathcal{O}}_{\eta_{c}}^{[1]}\left(^{1}S_{0}^{[1]}\right)\right\rangle \approx0.3\,{\rm GeV^{3}}$
is the dominant color singlet LDME of $\eta_{c}$ meson~\cite{Braaten:2002fi},
$\delta_{\eta_{c}}^{\mu}$ is the relative momentum of the $c$ and
$\bar{c}$ in the $\eta_{c}$ state, and $i,j$ are the color indices.
For light mesons, the quark mass is considered as a small parameter
whose contribution is controlled by higher twist corrections, so disregarding
mass term and using $\delta_{\eta_{c}}^{\mu}=\left(z-\frac{1}{2}\right)p_{\eta_{c}}^{\mu}$,
the projector~(\ref{eq:NRQCDProjector}) becomes proportional to
the leading twist projector 
\begin{equation}
\hat{\mathcal{P}}^{({\rm leading\,\,twist})}=\frac{1}{4}\hat{p}_{\eta_{c}}\gamma_{5}\otimes\frac{\delta_{ij}}{\sqrt{N_{c}}},
\end{equation}
whose structure could be deduced directly from~(\ref{eq:DAdefEta}). 

The interaction of the partonic ensemble with the target ($t$-channel
exchanges) in the leading twist may be described by the chiral even
gluon GPDs, which are defined as~\cite{Diehl:2003ny,DVMPcc1} 
\begin{align}
F^{g}\left(x,\xi,t\right) & =\frac{1}{\bar{P}^{+}}\int\frac{dz}{2\pi}\,e^{ix\bar{P}^{+}}\left\langle P'\left|G^{+\mu\,a}\left(-\frac{z}{2}n\right)\mathcal{L}\left(-\frac{z}{2},\,\frac{z}{2}\right)G_{\,\,\mu}^{+\,a}\left(\frac{z}{2}n\right)\right|P\right\rangle =\label{eq:defF}\\
 & =\left(\bar{U}\left(P'\right)\gamma_{+}U\left(P\right)H^{g}\left(x,\xi,t\right)+\bar{U}\left(P'\right)\frac{i\sigma^{+\alpha}\Delta_{\alpha}}{2m_{N}}U\left(P\right)E^{g}\left(x,\xi,t\right)\right),\nonumber \\
\tilde{F}^{g}\left(x,\xi,t\right) & =\frac{-i}{\bar{P}^{+}}\int\frac{dz}{2\pi}\,e^{ix\bar{P}^{+}}\left\langle P'\left|G^{+\mu\,a}\left(-\frac{z}{2}n\right)\mathcal{L}\left(-\frac{z}{2},\,\frac{z}{2}\right)\tilde{G}_{\,\,\mu}^{+\,a}\left(\frac{z}{2}n\right)\right|P\right\rangle =\label{eq:defFTilde}\\
 & =\left(\bar{U}\left(P'\right)\gamma_{+}\gamma_{5}U\left(P\right)\tilde{H}^{g}\left(x,\xi,t\right)+\bar{U}\left(P'\right)\frac{\Delta^{+}\gamma_{5}}{2m_{N}}U\left(P\right)\tilde{E}^{g}\left(x,\xi,t\right)\right).\nonumber \\
 & \tilde{G}^{\mu\nu,\,a}\equiv\frac{1}{2}\varepsilon^{\mu\nu\alpha\beta}G_{\alpha\beta}^{a},\quad\mathcal{L}\left(-\frac{z}{2},\,\frac{z}{2}\right)\equiv\mathcal{P}{\rm exp}\left(i\int_{-z/2}^{z/2}d\zeta\,A^{+}\left(\zeta\right)\right).
\end{align}
where $U,\bar{U}$ are the spinors of the incoming and outgoing proton.
The two-point gluon operators in~(\ref{eq:defF},~\ref{eq:defFTilde})
may be simplified in the light-cone gauge $A^{+}=0$ as 
\begin{align}
 & G^{+\mu_{\perp}\,a}\left(z_{1}\right)G_{\,\,\mu_{\perp}}^{+\,a}\left(z_{2}\right)=g_{\mu\nu}^{\perp}\left(\partial^{+}A^{\mu_{\perp},a}(z_{1})\right)\left(\partial^{+}A^{\nu_{\perp}a}(z_{2})\right),\\
 & G^{+\mu_{\perp}\,a}\left(z_{1}\right)\tilde{G}_{\,\,\mu_{\perp}}^{+\,a}\left(z_{2}\right)=G^{+\mu_{\perp}\,a}\left(z_{1}\right)\tilde{G}_{-\mu_{\perp}}^{\,a}\left(z_{2}\right)=\frac{1}{2}\varepsilon_{-\mu_{\perp}\alpha\nu}G^{+\mu_{\perp}\,a}\left(z_{1}\right)G^{\alpha\nu,\,a}\left(z_{2}\right)=\\
 & =\varepsilon_{-\mu_{\perp}+\nu_{\perp}}G^{+\mu_{\perp}\,a}\left(z_{1}\right)G^{+\nu_{\perp},\,a}\left(z_{2}\right)=\varepsilon_{\mu\nu}^{\perp}G^{+\mu_{\perp}\,a}\left(z_{1}\right)G^{+\nu_{\perp},\,a}\left(z_{2}\right)=\varepsilon_{\mu\nu}^{\perp}\left(\partial^{+}A^{\mu,a}(z_{1})\right)\left(\partial^{+}A^{\nu,\,a}(z_{2})\right).\nonumber 
\end{align}
After integration over $z$ in~(\ref{eq:defF},~\ref{eq:defFTilde}),
we effectively switch to the momentum space, where the derivatives
$\partial_{z_{1}}^{+},\,\partial_{z_{2}}^{+}$ reduce to the multiplicative
factors $k_{1,2}^{+}\sim\left(x\pm\xi\right)\bar{P}^{+}$, so the
Eqs.~(\ref{eq:defF},~\ref{eq:defFTilde}) can be rewritten as~\cite{DVMPcc1}
\begin{align}
\frac{1}{\bar{P}^{+}}\int\frac{dz}{2\pi}\,e^{ix\bar{P}^{+}}\left.\frac{\frac{}{}}{}\left\langle P'\left|A_{\mu}^{a}\left(-\frac{z}{2}n\right)A_{\nu}^{b}\left(\frac{z}{2}n\right)\right|P\right\rangle \right|_{A^{+}=0\,{\rm gauge}} & =\frac{\delta^{ab}}{N_{c}^{2}-1}\left(\frac{-g_{\mu\nu}^{\perp}F^{g}\left(x,\xi,t\right)-\varepsilon_{\mu\nu}^{\perp}\tilde{F}^{g}\left(x,\xi,t\right)}{2\,\left(x-\xi+i0\right)\left(x+\xi-i0\right)}\right).\label{eq:defF-1}
\end{align}
The Eq.~(\ref{eq:defF-1}) implies that in order to get the coefficient
functions $C_{\gamma\eta_{c}}$ and $\tilde{C}_{\gamma\eta_{c}}$,
we have to convolute the Lorentz indices of $t$-channel gluons in
diagrams of Figure~\ref{fig:Photoproduction-A} with $g_{\mu\nu}^{\perp}$
and $\varepsilon_{\mu\nu}^{\perp}$ respectively. In order to define
proper deformation of the integration contour near the poles of the
amplitude, we follow~\cite{DVMPcc1} and assume that the variable
$\xi$ in denominator should be replaced as $\xi\to\xi-i0$. 

\begin{figure}
\includegraphics[scale=0.6]{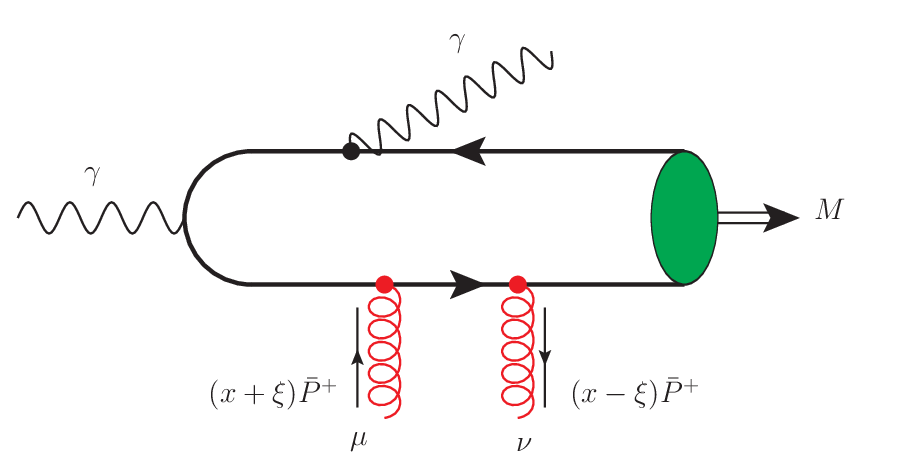}\includegraphics[scale=0.6]{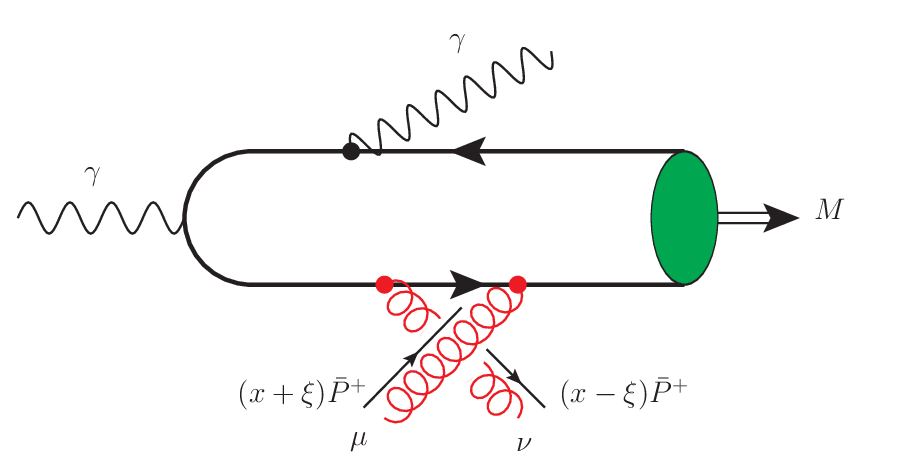}

\caption{\protect\label{fig:Photoproduction-Permute}An illustration of the
diagrams with direct and permuted $t$-channel gluons, which are related
to each other by inversion of sign in front of light-cone fraction
$x\leftrightarrow-x$, and permutation of the Lorentz indices $\mu\leftrightarrow\nu$.}
\end{figure}

The remaining steps in evaluation of the partonic amplitudes in Figure~\ref{fig:Photoproduction-A}
are straightforward and were performed using FeynCalc package for
Mathematica. The final result of this evaluation is
\begin{align}
c\left(x,\,\xi,\,z\right) & =\mathcal{C}\left(x,\,\xi,\,z\right)+\mathcal{C}\left(x,\,\xi,\,1-z\right)+\mathcal{C}\left(-x,\,\xi,\,z\right)+\mathcal{C}\left(-x,\,\xi,\,1-z\right),\label{eq:C}\\
\tilde{c}\left(x,\,\xi,\,z\right) & =\tilde{\mathcal{C}}\left(x,\,\xi,\,z\right)+\tilde{\mathcal{C}}\left(x,\,\xi,\,1-z\right)-\tilde{\mathcal{C}}\left(-x,\,\xi,\,z\right)-\tilde{\mathcal{C}}\left(-x,\,\xi,\,1-z\right),\label{eq:CTilde}
\end{align}
where the first terms $\mathcal{C}\left(x,\,\xi,\,z\right)$ and $\mathcal{\tilde{C}}\left(x,\,\xi,\,z\right)$
in the right-hand sides of~(\ref{eq:C},~\ref{eq:CTilde}) are the
contributions of the six diagrams shown explicitly in the Figure~$~\ref{fig:Photoproduction-A}$,
the second terms (with $z\to1-z$ substitution) were obtained from
charge conjugated diagrams with inverted direction of quark lines,
and the last two terms (with $x\to-x$ substitution) correspond to
diagrams with permuted gluons in the $t$-channel, as shown in the
Figure~\ref{fig:Photoproduction-Permute}. The structure of the right-hand
side of~(\ref{eq:C},~\ref{eq:CTilde}) implies that as a function
of $x$, the function $c$ is even, and $\tilde{c}$ is odd. Explicitly,
the components $\mathcal{C}^{(++)}\left(x,\,\xi,\,z\right)$ and $\tilde{\mathcal{C}}^{(++)}\left(x,\,\xi,\,z\right)$
are given by

\begin{align}
\mathcal{C}^{(++)}\left(x,\,\xi,\,z\right)= & \frac{\bar{\alpha}\mathfrak{C}_{\eta_{c}}}{4r^{2}\left(x-\xi+i0\right)\left(x+\xi-i0\right)}\times\label{eq:CF}\\
\times & \left[\frac{2\alpha z\left(2\alpha z+(\bar{\alpha}\bar{z}-\alpha)(\zeta+1)\right)-\frac{\alpha\left(4z^{3}+2z^{2}(\zeta+1)-z(4\zeta+5)+1\right)-\left(1-4z^{2}\right)\bar{z}}{r^{2}}+\frac{\left(1-4z^{2}\right)\bar{z}}{r^{4}}}{\left(\alpha\bar{z}\bar{\zeta}-\frac{4z^{2}-8z+3}{2r^{2}}-i0\right)\left(\alpha z(\zeta+1)+\frac{1-4z^{2}}{2r^{2}}-i0\right)\left((\zeta+1)+z(\alpha\,\bar{\zeta}-2)-\frac{1+4z\bar{z}}{2r^{2}}+i0\right)}\right.+\nonumber \\
 & +\frac{z\left(4\alpha\bar{\alpha}z-\alpha(\zeta+1)+2\bar{z}\right)-\frac{z}{r^{2}}(\zeta+1-6\alpha z+4z)-\frac{2z^{2}}{r^{4}}}{\left(\bar{z}-\frac{1+4z\bar{z}}{4r^{2}}+i0\right)\left(\zeta+1-z(2-\alpha\bar{\zeta})-\frac{1+4z\bar{z}}{2r^{2}}+i0\right)\left(\alpha z(\zeta+1)+\frac{1-4z^{2}}{2r^{2}}-i0\right)}\nonumber \\
 & +\frac{4\alpha z\left((4\alpha-2)z+\bar{\zeta}\right)-\frac{1}{r^{2}}\left(-8\bar{\alpha}z^{3}+16\alpha z^{2}+2\bar{\alpha}z-\bar{\zeta}\left(1-4z^{2}\right)\right)-\frac{2z\left(1-4z^{2}\right)}{r^{4}}}{4\left(\bar{z}-\frac{1+4z\bar{z}}{4r^{2}}+i0\right)\left(\alpha z+\frac{1-4z^{2}}{4r^{2}}-i0\right)\left(\alpha z\bar{\zeta}+\frac{1-4z^{2}}{2r^{2}}-i0\right)}\nonumber \\
 & -\frac{4\alpha^{2}z\bar{z}-\frac{1}{r^{2}}\left((8\alpha-4)z^{3}-4z^{2}(3\alpha+\zeta-1)+z(2\alpha+4\zeta+1)+\zeta+1\right)-\frac{2z\left(1+4z\bar{z}\right)}{r^{4}}}{4\left(\bar{\alpha}\bar{z}+\frac{1+4z\bar{z}}{4r^{2}}-i0\right)\left(\alpha z+\frac{1-4z^{2}}{4r^{2}}-i0\right)\left(\alpha z\bar{\zeta}+\frac{1-4z^{2}}{2r^{2}}-i0\right)}\nonumber \\
 & -\frac{\left(2\alpha\bar{\alpha}\bar{z}(4z-\bar{\zeta})+(2z-\bar{\zeta})^{2}\right)+\frac{4\alpha z^{3}+8z^{2}-20\alpha z^{2}+17\alpha z-2\bar{\alpha}\bar{z}\zeta-10z+2}{r^{2}}-\frac{4z^{3}-12z^{2}+7z+2}{r^{4}}}{2\left(\bar{\alpha}\bar{z}+\frac{1+4z\bar{z}}{4r^{2}}-i0\right)\left(\alpha z\bar{\zeta}+\frac{1-4z^{2}}{2r^{2}}-i0\right)\left((\alpha\bar{z}-1)\bar{\zeta}+2z-\frac{1+4z\bar{z}}{2r^{2}}+i0\right)}\nonumber \\
 & +\left.\frac{-4\bar{z}\left(4\alpha\bar{\alpha}\bar{z}-\bar{\alpha}\bar{\zeta}+2z\right)+\frac{2\alpha\bar{z}\left(4z^{2}+4z-9\right)-16z^{2}\bar{z}-\zeta(1+4z\bar{z})+4\bar{z}(z+5)+1}{r^{2}}+\frac{2\bar{z}(1+4z\bar{z})}{r^{4}}}{8\left(z-\frac{1+4z\bar{z}}{4r^{2}}+i0\right)\left(\bar{\alpha}\bar{z}+\frac{1+4z\bar{z}}{4r^{2}}-i0\right)\left((\alpha\bar{z}-1)\bar{\zeta}+2z-\frac{1+4z\bar{z}}{2r^{2}}+i0\right)}\right]\nonumber 
\end{align}
\begin{align}
\tilde{\mathcal{C}}^{(++)}\left(x,\,\xi,\,z\right)= & \frac{\bar{\alpha}\mathfrak{C}_{\eta_{c}}}{4r^{2}\left(x-\xi+i0\right)\left(x+\xi-i0\right)}\times\label{eq:CFT}\\
\times & \left[\frac{2\alpha z(-\alpha z\bar{\zeta}-3z\zeta+z+\zeta(\zeta+1))-\left(\frac{\alpha\left(12z^{3}+2z^{2}(\zeta-7)-3z-1\right)-4z^{3}+4z^{2}+z+1}{r^{2}}\right)-\left(\frac{-12z^{3}+12z^{2}+3z+1}{r^{4}}\right)}{\left(\alpha\bar{z}\bar{\zeta}-\frac{4z^{2}-8z+3}{2r^{2}}-i0\right)\left(\alpha z(\zeta+1)+\frac{1-4z^{2}}{2r^{2}}-i0\right)\left((\zeta+1)+z(\alpha\,\bar{\zeta}-2)-\frac{1+4z\bar{z}}{2r^{2}}+i0\right)}\right.\nonumber \\
 & +\frac{z\left(4\bar{z}+2\alpha(1-2z)(\zeta+1)\right)-\frac{z}{r^{2}}\left(\alpha\left(8z^{2}-4z-2\right)+4z^{2}(\zeta-3)+16z-3\zeta+1\right)+\frac{z}{r^{4}}\left(8z^{2}-4z-2\right)}{2\left(\bar{z}-\frac{1+4z\bar{z}}{4r^{2}}+i0\right)\left(\zeta+1-z(2-\alpha\bar{\zeta})-\frac{1+4z\bar{z}}{2r^{2}}+i0\right)\left(\alpha z(\zeta+1)+\frac{1-4z^{2}}{2r^{2}}-i0\right)}\nonumber \\
 & +\frac{4\alpha z((2z-1)\zeta+1)+\frac{4z\bar{z}(1+4z)-2\alpha z\left(1-4z^{2}\right)+\bar{\zeta}+2\zeta z\left(1+4z\bar{z}\right)}{r^{2}}+\frac{2z\left(1-4z^{2}\right)}{r^{4}}}{8\left(\bar{z}-\frac{1+4z\bar{z}}{4r^{2}}+i0\right)\left(\alpha z+\frac{1-4z^{2}}{4r^{2}}-i0\right)\left(\alpha z\bar{\zeta}+\frac{1-4z^{2}}{2r^{2}}-i0\right)}\nonumber \\
 & +\frac{4\alpha z\bar{z}(\bar{\alpha}+\zeta)+\frac{\alpha\left(4z\bar{z}-2\right)+\bar{z}\left(1+4z\bar{z}\right)+1}{r^{2}}+\frac{2}{r^{4}}}{4\left(\bar{\alpha}\bar{z}+\frac{1+4z\bar{z}}{4r^{2}}-i0\right)\left(\alpha z+\frac{1-4z^{2}}{4r^{2}}-i0\right)\left(\alpha z\bar{\zeta}+\frac{1-4z^{2}}{2r^{2}}-i0\right)}\nonumber \\
 & +\frac{\left(2\alpha^{2}\bar{z}\left(1-2z\right)\bar{\zeta}+2\alpha\bar{z}\bar{\zeta}(4z+\zeta-2)+(2z-\bar{\zeta})^{2}\right)}{2\left(\bar{\alpha}\bar{z}+\frac{1+4z\bar{z}}{4r^{2}}-i0\right)\left(\alpha z\bar{\zeta}+\frac{1-4z^{2}}{2r^{2}}-i0\right)\left((\alpha\bar{z}-1)\bar{\zeta}+2z-\frac{1+4z\bar{z}}{2r^{2}}+i0\right)}\nonumber \\
 & +\frac{1}{r^{2}}\frac{-\alpha r^{2}\left(12z^{3}-16z^{2}-z+4-2\zeta\bar{z}(1-2z)\right)-2r^{2}\bar{z}\left(4z^{2}+\zeta-2\right)+\left(2-3z\right)\left(1+4z\bar{z}\right)}{2\left(\bar{\alpha}\bar{z}+\frac{1+4z\bar{z}}{4r^{2}}-i0\right)\left(\alpha z\bar{\zeta}+\frac{1-4z^{2}}{2r^{2}}-i0\right)\left((\alpha\bar{z}-1)\bar{\zeta}+2z-\frac{1+4z\bar{z}}{2r^{2}}+i0\right)}\nonumber \\
 & +\left.\frac{4\bar{z}(2\alpha z(\zeta+1)-\bar{\alpha}\bar{\zeta}-2z)+\frac{-2\alpha\bar{z}\left(1+4z\bar{z}\right)+\bar{\zeta}(8z^{3}-20z^{2}+10z+1)-16z\bar{z}}{r^{2}}+\frac{2\bar{z}\left(1+4z\bar{z}\right)}{r^{4}}}{8\left(z-\frac{1+4z\bar{z}}{4r^{2}}+i0\right)\left(\bar{\alpha}\bar{z}+\frac{1+4z\bar{z}}{4r^{2}}-i0\right)\left((\alpha\bar{z}-1)\bar{\zeta}+2z-\frac{1+4z\bar{z}}{2r^{2}}+i0\right)}\right]\nonumber 
\end{align}
where $r=M_{\gamma\eta_{c}}/M_{\eta_{c}}\approx\sqrt{2\xi s}/2m_{c}$,
$\alpha\equiv\alpha_{\eta_{c}}$, $\bar{\alpha}\equiv1-\alpha_{\eta_{c}}$,
$\bar{z}\equiv1-z$, $\zeta=x/\xi$, the contour deformation prescription
was derived from the conventional $m_{c}^{2}\to m_{c}^{2}-i0$ prescription
in denominators of Feynman propagators, and the constant $\mathfrak{C}_{\eta_{c}}$
was defined earlier in~(\ref{eq:CDef}). The expressions for the
helicity flip components $\mathcal{C}^{(+-)},\,\mathcal{\tilde{C}}^{(+-)}$
have a similar structure, though have additional $\sim\mathcal{O}\left(1/r^{2}\right)$
suppression due to cancellation of the leading order term in the numerator.
In the limit $z_{1}=z_{2}=1/2$ these expressions reduce to~(\ref{eq:Monome}-\ref{eq:Monome-1})
from the main text. 

In our expressions~(\ref{eq:CF},\ref{eq:CFT}) we ordered the terms
by their behavior in the limit $r\to\infty$. This limit corresponds
to the high energy limit ($s\to\infty$) at fixed heavy quarkonium
mass $M_{\eta_{c}}$, as well as to the limit of negligibly small
meson mass at fixed invariant mass $M_{\gamma\eta_{c}}={\rm const}$
(so that $r=M_{\gamma\eta_{c}}/M_{\eta_{c}}\to\infty$). We may check
that in this limit for $C_{\gamma\eta_{c}}^{(++)}$ we may recover
the Eq\@. (A3) from~~\cite{Nabeebaccus:2023}, 
\begin{align}
 & \lim_{r\to\infty}C_{\gamma\eta_{c}}^{(++)}\left(x,\,\xi,\,z\right)\sim\frac{\left(1-\alpha/2\right)\left(x^{2}-\xi^{2}\right)}{\left(x+\xi-i0\right)^{2}\left(x-\xi+i0\right)^{2}}\times\label{eq:CInfty}\\
 & \times\frac{(1-\alpha)\left(\left(x^{2}-\xi^{2}\right)^{2}(1-2z\bar{z})+8\xi^{2}x^{2}z\bar{z}\right)+\left(x^{4}-\xi^{4}\right)z\bar{z}\left(1+\bar{\alpha}^{2}\right)}{\left(\bar{z}(x+\xi)-\bar{\alpha}z(x-\xi)-i0\right)\left(z(x-\xi)+\bar{\alpha}\bar{z}(x+\xi)-i0\right)\left(\bar{z}(x-\xi)+\bar{\alpha}z(x+\xi)-i0\right)\left(z(x+\xi)-\bar{\alpha}\bar{z}(x-\xi)-i0\right)}.\nonumber 
\end{align}
where we switched to conventional notations in terms of $x,\,\xi$.
As was explained in~\cite{Nabeebaccus:2023}, the coefficient function
~(\ref{eq:CInfty}) after convolution with gluon GPDs and meson LCDAs
might lead to the singular amplitudes when several singularities overlap
and start pinching the contour. Typically, this occurs when one of
the $t$-channel gluons is soft ($x=\pm\xi$) and interacts with a
soft quark from the final meson ($z$ or $\bar{z}=0$). From the structure
of the denominators in~(\ref{eq:CF},\ref{eq:CFT}) we can see that
in case of the massive theory with finite $r\not=\infty$ this no
longer happens, since $\sim\mathcal{O}\left(1/r^{2}\right)$ corrections
in denominators shift the position of the poles, thus avoiding overlaps.
For some terms the constants in front of $\sim\mathcal{O}\left(1/r^{2}\right)$
terms vanish in the point $z=1/2$, thus leading to an overlapping
poles at $x=\pm\xi$ . We checked explicitly that in all such points
these singularities are located on the same side of the integration
contour and do not pinch it. For this reason, the integration near
such second-order pole does not lead to physical singularity and may
be performed using the conventional rule~(\ref{eq:DoublePole}). 

In order to estimate numerically the role of the finite width of the
$\eta_{c}$ wave function, we calculated the cross-section using convolution
with normalized to unity light-cone DA from~~\cite{Dosch:1996ss}
. We found that this changes the cross-sections not more than 20 per
cent compared to NRQCD approach~(\ref{eq:F},\ref{eq:PhiHat}), on
par with expected contributions due to higher order $\mathcal{O}\left(\alpha_{s}(m_{c})\right)$
corrections.

 \end{document}